\PassOptionsToPackage{prologue,table,x11names,dvipsnames}{xcolor}
\documentclass[sigconf]{acmart}

\AtBeginDocument{%
  }

\setcopyright{acmlicensed}
\copyrightyear{2025}
\acmYear{2025}
\setcopyright{acmlicensed}\acmConference[UIST '25]{The 38th Annual ACM Symposium on User Interface Software and Technology}{September 28-October 1, 2025}{Busan, Republic of Korea}
\acmBooktitle{The 38th Annual ACM Symposium on User Interface Software and Technology (UIST '25), September 28-October 1, 2025, Busan, Republic of Korea}
\acmDOI{10.1145/3746059.3747711}
\acmISBN{979-8-4007-2037-6/2025/09}

\usepackage{makecell}
\usepackage{main,acmart-taps}
\usepackage{circledsteps}

\usepackage{amssymb}% http://ctan.org/pkg/amssymb
\usepackage{pifont}% http://ctan.org/pkg/pifont
\newcommand{\cmark}{\ding{51}}%
\newcommand{\xmark}{\ding{55}}%
\usepackage{tikz}
\usepackage{arydshln}

\newcommand{\new}[1]{\textcolor{black}{#1}}
\newcommand{\remove}[1]{\textcolor{red}{}
}

\begin{document}

\title{Knoll: Creating a Knowledge Ecosystem\\ for Large Language Models}
\author{Dora Zhao}
\affiliation{%
  \institution{Stanford University}
  \city{Stanford}
  \country{USA}
}
\email{dorothyz@stanford.edu}

\author{Diyi Yang}
\affiliation{%
  \institution{Stanford University}
  \city{Stanford}
  \country{USA}
}
\email{diyiy@stanford.edu}

\author{Michael S. Bernstein}
\affiliation{%
  \institution{Stanford University}
  \city{Stanford}
  \country{USA}
}
\email{msb@cs.stanford.edu}

\renewcommand{\shortauthors}{Zhao et al.}


\begin{abstract}
% Large language models are designed to encode general purpose knowledge about the world from Internet data. Yet, there is a wealth of information that falls outside of this scope --- ranging from personal preferences to organizational policies, from community-specific advice to the most up-to-date news --- that users may want their models to have access to, but is currently unavailable. In this paper, we introduce software infrastructure that enables end users to create, curate, and configure custom knowledge modules that are utilized by language models such as ChatGPT and Claude. Our infrastructure, Knoll, allows users to create modules by clipping content from the web or authoring shared documents on Google Docs and GitHub, add modules that others have made, and then automatically inserts the relevant knowledge when users interact with an LLM. We conducted a public deployment of Knoll reaching over 200 users who employed the system for a diverse set of tasks including personalized recommendations, advice-seeking, and writing assistance. Finally, we validate that using Knoll improves the quality of generated responses. 
Large language models are designed to encode general purpose knowledge about the world from Internet data. Yet, a wealth of information falls outside this scope --- ranging from personal preferences to organizational policies, from community-specific advice to up-to-date news --- that users want models to access but remains unavailable. In this paper, we propose a knowledge ecosystem in which end-users can create, curate, and configure custom knowledge modules that are utilized by language models, such as ChatGPT and Claude. To support this vision, we introduce Knoll, a software infrastructure that allows users to make modules by clipping content from the web or authoring shared documents on Google Docs and GitHub, add modules that others have made, and rely on the system to insert relevant knowledge when interacting with an LLM. We conduct a public deployment of Knoll reaching over 200 users who employed the system for a diverse set of tasks including personalized recommendations, advice-seeking, and writing assistance. \new{Knoll improves the quality of generated responses with participants preferring responses generated with Knoll over baseline \texttt{GPT-4o} responses for 81.5\% of the queries when external knowledge is needed.} 
\end{abstract}




 \begin{teaserfigure}
    \centering
    \includegraphics[width=\linewidth]{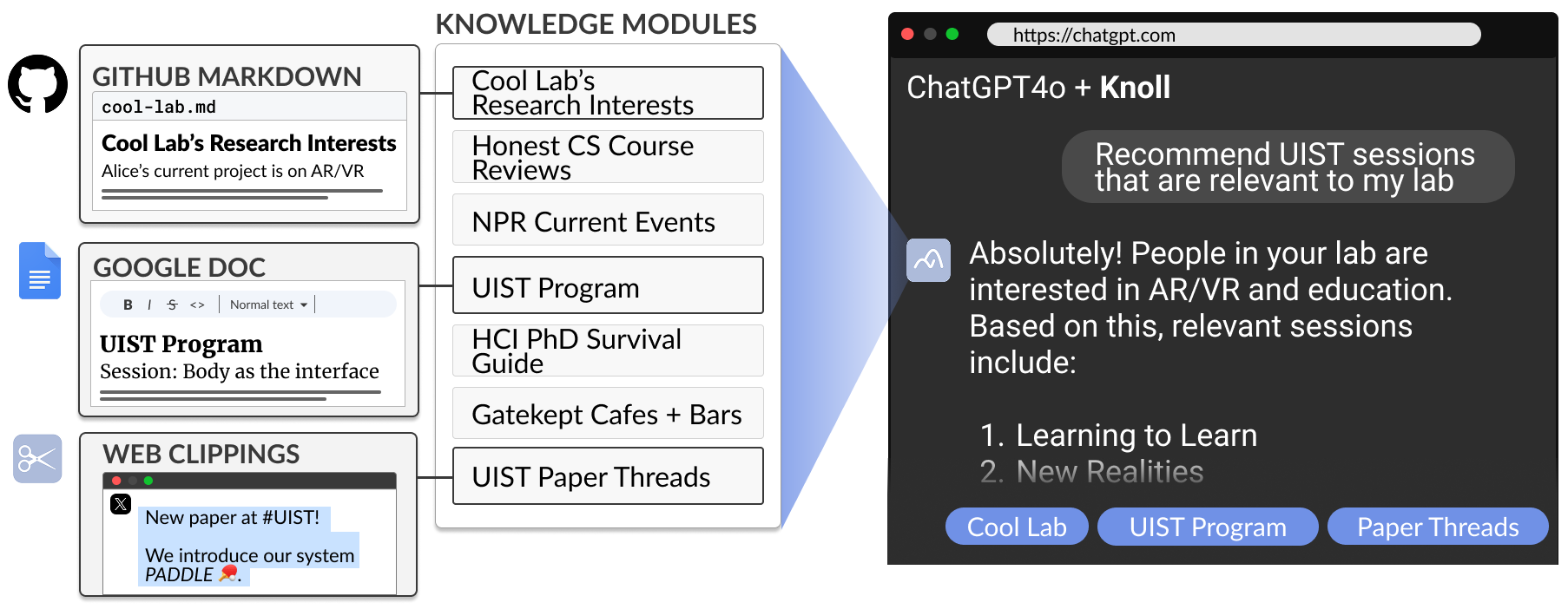}
    \Description[Diagram showing how Knoll works]{Diagram showing how users create and share modular knowledge, which can then be injected into prompts for large language models. On the left, we show how users can author, clip, and share modules. On the right, we show how LLMs automatically incorporate the selected modules into responses on the web client.}
    \caption{Knoll enables an ecosystem of \textit{knowledge modules} that enable end users to provide their large language model access to specific, local knowledge. Within this ecosystem, users can author new modules, share these modules with others, and incorporate modules created by the community (left). When querying an LLM, the model automatically incorporates relevant knowledge from the selected modules (right).}
    \label{fig:teaser}
\end{teaserfigure}
\maketitle

% \section{Introduction}
% \input{Sections/Intro} 

\section{Introduction}
There is much that large language models (LLMs) do not know. LLMs embed a general understanding of the world from their pre-training on internet data~\cite{petroni2019language,roberts2020much}, enabling them to solve challenging math problems~\cite{liu2024deepseek}, write essays~\cite{gomez2023confederacy}, and generate code~\cite{li2022competition,jimenez2024swe}. But much of the knowledge that we leverage in our everyday lives is not in LLMs' pretraining data~\cite{haraway2013situated}. As a result, LLMs still struggle with tasks that require local knowledge --- such as applying community norms within a specific group (``How does my UIST paper introduction match or differ from my lab's style and advice?'') or handling questions that require localized domain expertise (``How should I prepare for my qualifying exams in Stanford's HCI Group'') --- and end up producing overly generic or even hallucinated answers~\cite{ma2024evaluating,shuster2021retrieval,li2024enhancing}. This shortcoming highlights a gap between what models have access to in their training data and the abundance of knowledge that people make use of in their day-to-day lives that models are not yet privy to. 

Existing methods focus on integrating external knowledge into language models, but models cannot retrieve knowledge that has never been written down. In cases where knowledge is already recorded, for example an organization's internal documents, retrieval-based methods (e.g., RAG) allow models to access that knowledge and insert it into the model's context~\cite{lewis2020retrieval}. These methods can succeed when knowledge already exists in an easily accessible format, such as Wikipedia data or domain-specific corpora~\cite{xiong2024benchmarking,wiratunga2024cbr,kang2024biospark}. Likewise, commercial products, such as Open AI’s Custom GPT~\cite{customgpt} or Anthropic’s Claude Projects~\cite{claude}, allow users to upload existing files to create custom models tailored for a particular task.

However, these approaches fail to address the knowledge gap issue. Much of the information that individuals or communities want to use is diffuse --- fragmented across sources~\cite{bernstein2008information} or people~\cite{tsoukas1996firm,hutchins2000distributed} --- or never even written down~\cite{howells1996tacit,nonaka2009knowledge}. As a result, this knowledge remains largely inaccessible to retrieval-based methods. \new{Although there have been many technical solutions proposed for how to incorporate external knowledge into language models~\cite{mcp,lewis2020retrieval},} we are still left with fundamental questions around \textit{how} to empower individual end users and communities to fill these knowledge gaps.  

In this work, we propose an alternative vision in which end-users access and contribute to an \emph{ecosystem of knowledge modules} which are used to augment their LLMs with custom information. We envision an analogue to the Python Package Index or npm Registry, where you can import information for your ChatGPT or Claude client to access and contribute modules for yourself or for friends, colleagues, or the world to use. We define a ``knowledge module'' as a self-contained repository of information that an individual wants their LLM to access: for example, a Markdown file with project meeting notes, a private Google Document shared within the research lab with ``hot takes'' on writing UIST papers, or a clipping from social media explaining a new paper of interest. 

Our ecosystem consists of two activities:~(1) knowledge creation, where users can author modules intended for personal use or for sharing with the broader community, and~(2) knowledge consumption, where these modules are integrated when interacting with LLMs. These activities are not mutually exclusive; individuals can both create and consume modules, allowing the ecosystem to function at different scales, from a single person making use of personalized modules to a broader community where modules are shared and maintained across many users. \new{Importantly, this ecosystem provides a direct pathway for end-users to easily curate and integrate the knowledge they want to access into an LLM.}

We instantiate this idea in \emph{Knoll}, a browser extension that \new{allows users to make use of this ecosystem of knowledge modules when interacting with commercial LLM chatbots.\footnote{\new{The project page for Knoll can be found at \url{https://stanfordhci.github.io/knoll/}.}} With Knoll, users are able to specify what knowledge they want their model to have access to by importing existing modules or creating new modules via ``clipping'' any text they come across on the internet or directly adding documents.} Then, when a user queries a language model via the typical web interface, Knoll automatically selects which modules are useful and injects them into the prompt. 

To evaluate Knoll, we conduct a public deployment. During a two-month period, 203 users downloaded Knoll and sent over 1,500 queries with information added from Knoll knowledge modules. These users imported 30 unique modules and created nine new modules, spanning both personal use (e.g., final exam study guides for an advanced HCI course, ML / research engineer job postings) and community-oriented information (e.g., policies for a course at Stanford University, lab-specific paper writing norms). Using data collected from our public deployment, we first conduct a technical evaluation of Knoll using a subset of 100 queries to test whether adding knowledge modules improves the quality of generated responses. We find that annotators prefer responses generated with Knoll over \texttt{GPT-4o} responses for 81.5\% of the queries when external knowledge is needed. Knoll struggles in cases where external knowledge is related to the query but not necessary for generating a response, as the model may over-rely on the provided information. For example, when a user asked ChatGPT to fix the grammar in their essay, Knoll incorporated knowledge on the assignment's guidelines, leading the model to edit the essay's content. Second, we triangulate between log data (N=203), survey responses (N=17), and semi-structured interviews (N=8) to analyze usage patterns. We find that Knoll helps ameliorate data void issues with LLMs~\cite{boyd2018data}, empowering users to ask queries to LLMs they otherwise would not have due to hallucinations or a lack of relevant information. 

Taken together, this work introduces the concept of a modular knowledge ecosystem in which users can create, collaborate, and configure custom knowledge for their language model. To support this ecosystem, we introduce software infrastructure, Knoll, that supports this ecosystem, building tools that enable users to incorporate diffuse, localized information sources into LLMs thus filling existing knowledge gaps. We evaluate Knoll's ability to improve the quality of model generations and deploy it for real-world usage.

\section{Related Work}
    Knoll draws upon existing work on augmenting language models with external knowledge, end-user customization, personal information management, and language model interface design.

\subsection{Knowledge-Enhanced Language Models}
Large language models struggle with knowledge-intensive tasks~\cite{ji2023survey,kandpal2023large,mallen2023not}. To address this issue, prior works have proposed approaches to augment language models with externally stored knowledge~\cite{jiang2024towards,gururangan2020don}. One widely adopted approach is \emph{retrieval augmented generation (RAG)}. At a high-level, RAG-based approaches consist of retrieving relevant knowledge from an external vector database to use during generation. Other methods include leveraging smaller language models that have been fine-tuned on specialized corpora as ``plug-ins'' to larger general-purpose models~\cite{feng2024knowledge}. These approaches provide us with technical solutions for addressing the knowledge gap. However, they fail to wrestle with the underlying sociotechnical question of how this knowledge gets generated, who is able to generate it, and how this knowledge is maintained. As Kraft et al.~\cite{kraft2024knowledge} argue, existing approaches advance the problematic notion of ``the view from nowhere'' within AI, or the idea that knowledge is objective and neutral, when in fact these datasets are undoubtedly shaped by the politics of their creators~\cite{scheuerman2021datasets,santy2023nlpositionality}. 

In response, there has been interest in tailoring LLMs to reflect the knowledge of specific communities or individuals. 
One line of work has sought to align models to group preferences~\cite{wang2024social,he2024community,ramesh2024group,feng2024modular}. For example, Wang et al.~\cite{wang2024social} presented a framework for grounding LLM-based agents in the norms of different online groups using social signals found in existing chat history. Simultaneously, efforts to create personalized LLMs have leveraged a range of input --- including naturally occurring feedback in dialogue~\cite{don2024learning,petrak2023learning}, synthetically generated user data~\cite{singh2025fspo}, and user demonstrations~\cite{shaikhaligning} --- to align model outputs to user preferences. 

More closely related to our work, there has been interest in allowing users to define their own knowledge sources. These efforts have been integrated into existing commercial offerings. For example, OpenAI's Custom GPTs allows users to build their own versions of GPT by uploading relevant documents or writing tailored prompts~\cite{customgpt}. Similarly, Anthropic's Claude Projects is designed such that users can upload internal knowledge (e.g., documentation, notes) to personalize the model's outputs into a dedicated workspace~\cite{claude}. While we agree that users should have the ability to import their own knowledge sources when interacting with LLMs, these solutions compartmentalize knowledge into separate versions of the model that the user has to explicitly summon. Instead, localized knowledge should be weaved into the general model that the user interacts with --- integrated on-demand and without extra work from the user.

\subsection{End-User Customization}
Allowing end-user customization of technical systems has long been a goal in human-computer interaction. Prior works developed tools that allowed users to manipulate UIs via end-user programming or direct manipulation~\cite{bolin2005automation, nebeling2013crowdadapt,nebeling2016xdbrowser}. Advancements in large language model capabilities has made end-user customization more accessible: users can now specify interface changes in natural language~\cite{kim2022stylette}. This goal of end-user customization extends beyond web design. Another line of work has focused on end-user control over AI systems~\cite{fails2003interactive,amershi2014power}, proposing tools that empower users to train and update their own models~\cite{fails2003interactive,lam2023model}, correct model errors~\cite{lee2024clarify}, and curate their own evaluation datasets~\cite{kuo2024wikibench}. 

With the popularization of LLMs that are ``general-purpose'' by design, there continues to be a need to personalize these models. One approach has sought to employ algorithmic methods to personalize models from interaction data~\cite{tan2024personalized,wu2025aligning,don2024learning,shi2024life}. For example, ChatGPT's ``memory'' feature stores information learned about the user based on past messages, which are then used at generation time to tailor the model’s responses~\cite{memories}. A shortcoming with this approach, however, is the limited user autonomy. To address these issues, Memolet~\cite{yen2024memolet} and Memory Sandbox~\cite{huang2023memory} offer new affordances that allow users to directly manipulate memories and engage in more in-depth sense-making. However, memory is designed to captures a bank of personal user preferences and information (e.g., ``prefers short answers to essay responses,'' ``owns a bakery'') that is broadly applicable. Our work aims to go beyond individual preferences and integrate a broader range of localized knowledge into LLMs. 

\subsection{Personal Information Management}
Our work builds on existing literature that has examined how people manage their personal information, since this information is often the kind of knowledge that people feel is relevant for their language model. Prior work has shown how personal information is often stored in ``information scraps'' that never make their way into more traditional information management systems~\cite{bernstein2008information}. For example, people use Post-It notes to jot down a new idea or have an unnamed text file with miscellaneous content saved~\cite{fischel2018survey}. Flexibility both in contents and purpose, along with their lightweight nature, make information scraps useful~\cite{bernstein2008information}.

Given the diffuse nature of information scraps, how do people manage this information? Here, we focus on web-based resources. One popular method built into browsers is the bookmarking tool. Users can save sites that they visit and organize them into hierarchical folders to revisit later. While the native bookmarking function only saves the link, Hu and Lee~\cite{hu2022scrapbook} proposed Scrapbook, enabling users to create screenshot-based bookmarks for more efficient retrieval of digital resources. Similar to bookmarking, clipping, which allows users to save web content to an external storage system, has been explored extensively in academic research~\cite{kuznetsov2022fuse,rachatasumrit2021forsense,kittur2014standing,liu2022wigglite,kang2022threddy} and integrated in commercial products (e.g., Obsidian\footnote{https://obsidian.md/clipper}, Evernote\footnote{https://evernote.com/features/webclipper}). 

We draw upon existing work on personal information management when designing Knoll. From prior work, we know that users want their LLMs to have access to more contextualized information~\cite{yun2025generative}. However, as the literature on information scraps would suggest, the actual knowledge that may be relevant to an LLM is unlikely to be stored in a centralized location~\cite{bernstein2008information}. Thus, we present a way for users to capture this diffuse information in a format that is accessible to an LLM. 

\subsection{Human-LLM Interaction Design}
How can we improve user interactions with large language models? Prior work has shown that users struggle when prompting language models~\cite{zamfirescu2023johnny,subramonyam2024bridging,jiang2022promptmaker}. To bridge this gulf of execution, researchers have proposed systems that make prompt engineering more accessible. One approach has been to provide visual interfaces that allow users to explore different prompt options~\cite{arawjo2024chainforge,angert2023spellburst}. For example, Arawjo et al.'s~\cite{arawjo2024chainforge} ChainForge provides a visual programming environment for prompt engineering in which users can query different models and synthesize over model outputs. Other works have proposed new interface layers that shift the interaction from prompt editing to direct manipulation~\cite{masson2024directgpt,masson2024textoshop} or towards being more object-oriented in nature~\cite{kim2023cells,riche2025ai}. Parallel efforts have sought to support users in making sense of LLM outputs. For example, systems such as Graphologue~\cite{jiang2023graphologue} and Sensecape~\cite{suh2023sensecape} visualize model responses as interactive diagrams. Our work is complementary to this literature. Instead of focusing on the inputs and outputs to models, we intervene at the data layer, providing a way for end users to customize the external knowledge used by LLMs.

\section{Knoll}
\label{sec:knoll}
\renewcommand{\arraystretch}{1.25} % Default value: 1
\begin{table*}[]
    \centering
    \begin{tabular}{p{1.05in}>{\raggedright\arraybackslash}p{2.2in}p{0.7in}p{0.8in}p{0.7in}p{0.7in}}
    \toprule 
    & \textbf{Description} & \makecell{\textbf{Knowledge}\\\textbf{Creation}}
 & \makecell{\textbf{Collaborative}\\\textbf{Management}} & \makecell{\textbf{Automatic}\\\textbf{Integration}} & \makecell{\textbf{Platform}\\\textbf{Agnostic}}\\
    \midrule
    Prompting & Users can directly add relevant context in the query sent to the model. & \makecell{\xmark} & \makecell{\xmark} & \makecell{\xmark} & \makecell{\cmark} \\
    File Upload & Users can attach relevant files to each query. & \makecell{\xmark} & \makecell{\xmark} & \makecell{\xmark} & \makecell{\cmark}\\
RAG~\cite{lewis2020retrieval} & \new{A method for accessing and integrating information from an external data source when generating responses with a LLM.} & \makecell{\xmark} & \makecell{\xmark} & \makecell{\cmark} & \makecell{\cmark}\\ 
    Model Context\newline Protocol (MCP)~\cite{mcp} & \new{A standardized protocol that allows models to connect with external data sources and tools. While MCP is platform agnostic, not all commercial LLMs offer integrations with MCP servers.} & \makecell{\xmark} & \makecell{\xmark} & \makecell{\cmark} & \makecell{\cmark} \\
    LLMs + Web Search \cite{gptweb,claudeweb} & Language models can search the web for up-to-date information. While web knowledge is agnostic to the platform, not all commercial LLMs offer this service. & \makecell{\xmark} & \makecell{\xmark} & \makecell{\cmark} & \makecell{\cmark}\\
    GPT Memory~\cite{memories} & As the user interacts with ChatGPT, the system will learn information about the user (e.g., ``I am an HCI researcher) and their preferences (``I prefer bullet-point lists.''). & \makecell{\xmark} & \makecell{\xmark} & \makecell{\cmark} & \makecell{\xmark}\\
    Custom GPTs~\cite{customgpt} & Users can create custom versions of GPT that are tailored for particular tasks through crafting system prompts, uploading files, and adding access to tools. & \makecell{\xmark} & \makecell{\cmark} & \makecell{\xmark} & \makecell{\xmark} \\
    Claude Projects~\cite{claude} & Similar to Custom GPTs, projects are ``self-contained workspaces'' where users can upload relevant documents and create custom system prompts to tailor their conversations.  & \makecell{\xmark} & \makecell{\cmark} & \makecell{\xmark} & \makecell{\xmark}\\
    \rowcolor{lightgray!40} \textbf{Knoll} & Users can automatically integrate relevant external knowledge in the form of ``knowledge modules.'' Users can create modules by creating documents, clipping web content, or importing modules from other users. & \makecell{\cmark} & \makecell{\cmark} & \makecell{\cmark} & \makecell{\cmark}\\
    \bottomrule 
    \end{tabular}
    \Description[Comparison table of knowledge augmentation methods]{Comparison table summarizing the features of various knowledge augmentation methods for LLMs, including prompting, file upload, LLM with web search enabled, GPT memory, Custom GPTs, Claude Projects, and Knoll. Knoll is the only approach that supports all four dimensions: knowledge creation, collaborative management, automatic integration, and platform agnosticism.}
    \caption{Knoll offers four main benefits in comparison to existing methods. First, we help scaffold the knowledge module creation process. Second, users can collaboratively create and maintain modules as well as share modules with others. Third, the knowledge integration happens directly in the main chat interface. Rather than having to navigate to an external interface or manually upload the information, Knoll automatically selects and injects relevant knowledge. Finally, the knowledge stored in Knoll is platform agnostic, allowing users to have shared contexts across different LLM providers.}
    \label{tab:comparisons}
\end{table*}
\begin{figure}[b]
    \centering
    \includegraphics[width=\linewidth]{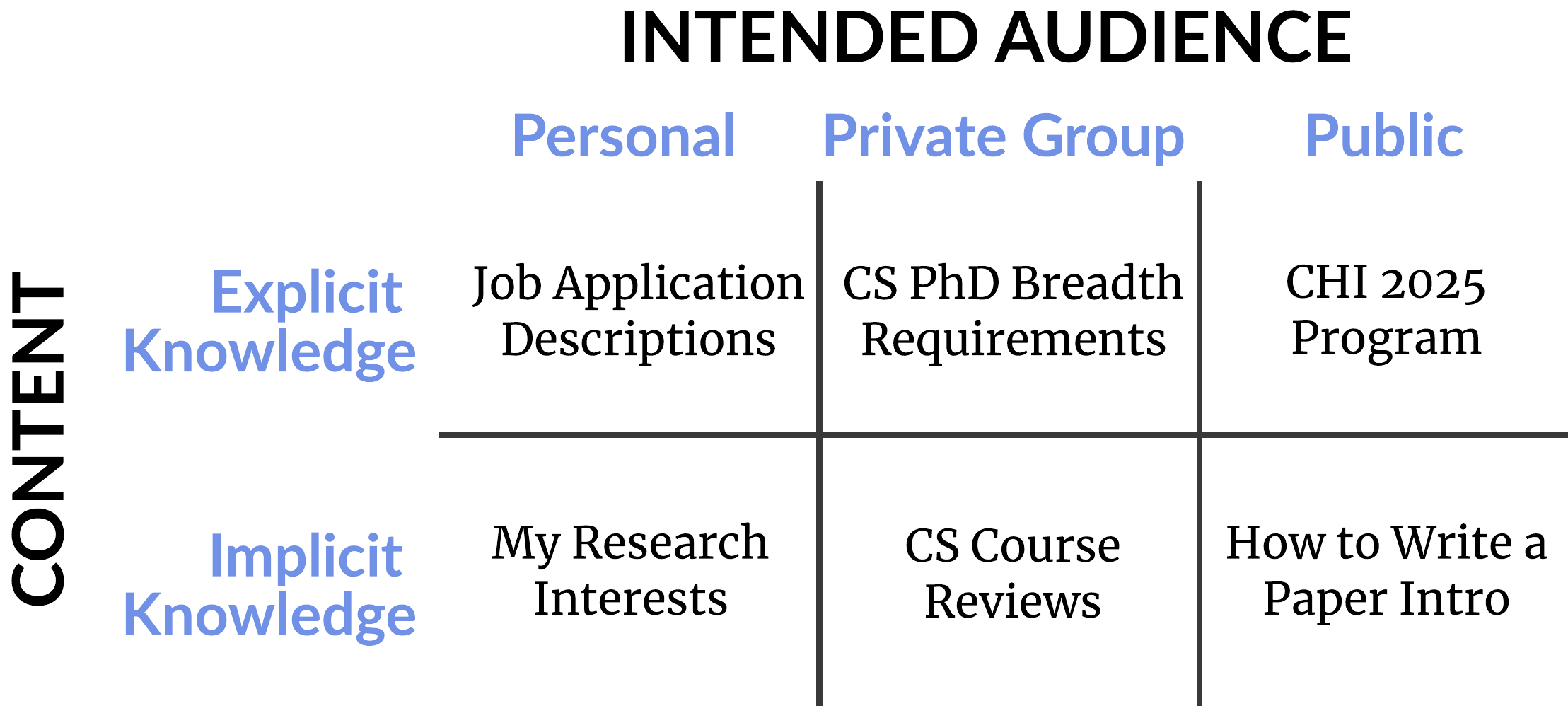}
    \Description[Matrix showing design space of knowledge modules]{Matrix illustrating the design space of knowledge modules along two dimensions: audience and knowledge type. The x-axis ranges from personal to public, and the y-axis ranges from explicit to implicit knowledge. Example modules include “CS PhD Breadth Requirements,” “Job Application Descriptions,” and “How to Write a Paper Intro.”}
    \caption{Knowledge modules can range in (1) their intended audience, including content that is personal, intended for a private group, or for the public and (2) the type of content, including more explicit, structured knowledge and implicit, unstructured knowledge. We provide examples of modules from our public deployment (Sec.~\ref{sec:field}) that span this design space.}
    \label{fig:design}
\end{figure}
Consider the ability to connect your language model to a kind of GitHub for localized knowledge --- one that contains both public and private repositories of knowledge that you can import. Knoll, named as a reference to ``knowledge'', provides software infrastructure that enables this integration of customized information when interacting with LLMs in the form of \emph{knowledge modules}. These modules are designed to contain any external information that an end-user wants to an LLM to use when generating responses. When relevant, these modules are inserted into queries sent to the LLM. As shown in Fig.~\ref{fig:design}, modules can differ in the type of content --- ranging from structured, declarative knowledge (e.g., policies, documentation) to unstructured, implicit knowledge (e.g., experiences, norms) --- and  intended audience. 

In this section, we detail the key design principles behind Knoll, comparing our infrastructure to existing methods for integrating customized knowledge into LLMs. We then provide an example usage scenario before discussing how users can create, curate, and configure modules with Knoll.

\subsection{Design Principles}
When creating Knoll, we emphasized four main design goals. 
\begin{enumerate}
    \item \textbf{Supporting Knowledge Creation}: First, we seek to support users in creating and managing knowledge sources. As shown in Table~\ref{tab:comparisons}, most existing methods assume the information that a user wants to consume is publicly available on the internet or already exists in a static and easily accessible format, such as internal style guides or software documentation, \new{and focus on the technical problem of how to best retrieve or incorporate this knowledge}. This assumption contradicts findings from prior work that surface how information is often scattered~\cite{bernstein2008information}, motivating a design that enables users to easily synthesize these fragmented sources. 
    \item \textbf{Fostering Collaborative Knowledge Management}: Second, knowledge management is a collaborative process~\cite{stahl2006group,kuznetsov2006motivations,ackerman2000reexamining}. Drawing from literature on peer production~\cite{benkler2015peer}, we expect that it is unlikely for a single user to have enough background knowledge or motivation to create all their modules. Instead, users should be able to work with others to author, contribute, and maintain this knowledge base.  
    \item \textbf{Allowing Automatic Knowledge Integration}: Existing methods either require users to input external knowledge for each conversation (e.g., copying into a prompt, uploading a file) or navigate to a separate workspace --- both of which are tedious. We can provide a more seamless interaction by supplying relevant knowledge as context to the model, all within the main chat window. 
    \item \textbf{Providing a Platform Agnostic Knowledge Source}: Following both normative arguments for more open and decentralized language model development~\cite{raffel2023building,kapoor2024position,tan2025medforge} and the more practical consideration that many users rely on multiple LLM services, we seek to provide infrastructure that is independent from platforms. Users maintain a consistent knowledge source across services and ultimately retain ownership over their knowledge modules rather than having it tied to one particular platform.  
\end{enumerate}

% As shown in Table~\ref{tab:comparisons}, existing methods for integrating knowledge tend to be laborious (e.g., including information in the prompt, uploading relevant files each time) or tied to a specific platform. In addition, almost all methods assume the knowledge that users want to add already exists in an easily accessible format. In contrast to existing offerings, Knoll supports users in creating and managing the information they want to add. Furthermore, knowledge is automatically integrated when relevant to the user's query, minimizing the need to create new interfaces, and can be used across different LLM providers. 

\subsection{Example Scenario}
Let us walk through an example scenario to demonstrate our goals for our knowledge module ecosystem. Vicky is entering her first year of a PhD program in computer science. She has questions about choosing an advisor at her university, so she decides to ask ChatGPT for advice. However, the model responds with generic information that can be found on the professors' websites. Later, as Vicky browses Knoll's Explore page, she finds a module that another student in the program has made with compiled advice and experiences. Vicky adds the module to her browser extension, and then navigates to ChatGPT where she prompts the model again. Now, ChatGPT incorporates student perspectives on the professors' advising style and lab culture when responding. 

Vicky wants to further personalize her model with information for first-years in the department. She creates a Google Doc with information that the department has sent her, which she uploads as a module called ``First-Year CS PhD Info.'' Vicky also uses the clipping feature to add other information she comes across, such as internal department fellowships and relevant courses. Now, she can query her ChatGPT for more tailored recommendations, such as ``which courses are most helpful for preparing for qualifying exams'' or ``what funding opportunities am I eligible to apply for and are related to my research.'' 

Since Vicky found her new module to be helpful, she wants to share it with others. She shares a link to the module with her cohort, so that they can add the ``First-Year CS PhD Info'' module to their Knoll extension via the link. They use the module for different tasks, including finding interesting courses, navigating department policies, and seeking advice on acclimating to the PhD program. Other cohort members add information about their first-year experiences to the module, collaboratively growing the knowledge base. 

\subsection{Interacting with Knoll}
\begin{figure*}
    \centering
    \includegraphics[width=\linewidth]{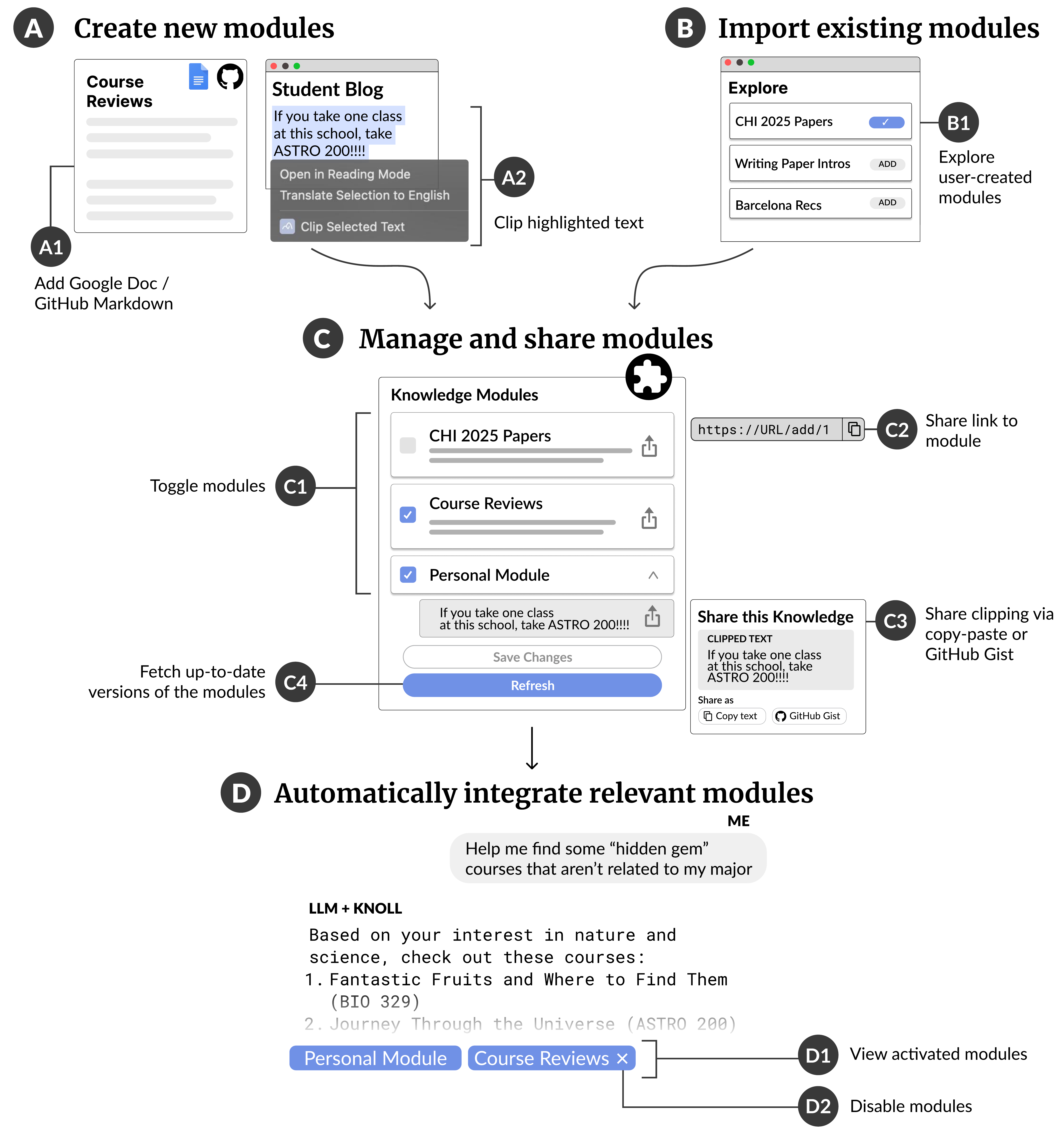}
    \Description[Four-part user interface overview of Knoll]{Four-part user interface overview of Knoll. (A) Users create modules by adding Google Docs, GitHub Markdown, or web clippings. (B) They import existing modules through a module gallery. (C) The extension lets users manage modules—toggle on/off, share, or refresh content. (D) When a query is submitted, Knoll displays which modules were used via chips at the bottom of the LLM response.}
    \caption{Knoll's interface allows users to create or add external knowledge and then automatically integrates this knowledge when relevant to the user's query. \myCircled{A} Users are able to create knowledge by adding Google Docs / Markdown files on Github or by clipping web content. \myCircled{B} They can also import existing modules that other users have created. Via the Knoll browser extension \myCircled{C}, users can toggle modules on-and-off and share modules. \myCircled{D} When a user submits a query to an LLM, Knoll automatically selects the external knowledge that is relevant and integrates it into the query using the pipeline we describe in Sec.~\ref{sec:router}. We display which modules were used as interface chips rendered at the bottom of the response.}
    \label{fig:system}
\end{figure*}

Next, we provide more details on how users interact with Knoll. Our infrastructure supports users in curating a custom knowledge base for LLMs through creating new modules or importing modules others have already contributed to the ecosystem. Users can rely on Knoll to automatically select modules relevant to their query and inspect what modules the model used when generating its response. 

To access Knoll, users download our browser extension, which is available on the Chrome Web Store.\footnote{\new{\url{https://chromewebstore.google.com/detail/knoll/fmboebkmcojlljnachnegpbikpnbanfc}}} They can use the browser extension to manage their knowledge modules. We also have an accompanying web application where users can explore and import modules that others have made.\footnote{\new{\url{https://knollapp.com/}}} By default, all users join Knoll under guest accounts; however, if they want to access certain privacy and sharing features, they must authenticate using GitHub or Google sign-in.\footnote{\new{While this work focuses on using a browser extension to integrate knowledge modules, there are alternative technical implementations. For example, we also have implemented an MCP server that integrates with Knoll: \url{https://github.com/StanfordHCI/knoll-mcp}.}}

\subsubsection{Creating New Knowledge Modules}
Users can create new knowledge modules by adding documents or clipping web content (Fig.~\ref{fig:system} \filledcircle{A}). Knoll integrates with collaborative tools, allowing users to convert Google Docs or Markdown files \new{on GitHub} \filledcircle{A1} into modules. Users begin by writing in either platform and, when ready, import the content by providing the document's URL. The system then retrieves the content via either Google Docs or GitHub API. To keep modules up-to-date, users can click the ``Refresh'' button~\filledcircle{C4} in their browser extension, which will then import the current version of the document.

By building on existing collaborative software, we inherit these platforms' fine-grained data access control functionalities. By changing the repository's sharing settings on GitHub or document visibility on Google, users maintain control over who can import their module and who can edit their module. With this functionality, users can also craft private modules that only they can access. 

Peer production systems need users to contribute content to be sustainable~\cite{cosley2007suggestbot,benkler2006commons}. The same principle applies to Knoll. To lower the barrier to module creation, we sought to enable people to capture knowledge while they are web browsing. As shown in~\filledcircle{A2}, Knoll includes a clipping feature that lets users add any online text to a module by highlighting it and selecting “Clip Selected Text” from the pop-up menu. By default, clipped content is saved to the user's \emph{Personal Module}, a private collection stored locally in the browser extension. Users can also manually edit this module or choose to save content directly to a Google Doc or GitHub Markdown file.

\subsubsection{Importing Existing Modules} 
Users can import modules that other Knoll users have created. On the \emph{Explore Page} \filledcircle{B1} of our web application, users browse our gallery or search for specific keywords to find modules to add to their extension. Users can inspect module details, which includes viewing the source content and example queries provided by the creator to illustrate the types of tasks the module can support. By clicking the ``add'' button, users import and activate the module, enabling the LLM to incorporate its knowledge in responses.

\subsubsection{Sharing Modules with the Community}
As part of the knowledge ecosystem, users can share modules that they have created with others in the community. Users can choose to contribute their module to the gallery on the \emph{Explore Page}. Alternatively, they can share their modules via a link that we generate \filledcircle{C2}. To retain data privacy, only users who have permissions to view the source document (e.g., Google Docs permissions or GitHub permissions, depending on where the document is hosted) are able to access shared modules. Finally, we provide two ways for users to share clipping added into their Personal Modules \filledcircle{C3}: copying the text to their clipboard or automatically creating a GitHub Gist.  

\subsubsection{Seamlessly Interacting with LLMs}
Users can interact with LLMs on existing web platforms (https://chatgpt.com, https://claude.ai) just as they normally would. When a query is sent, Knoll automatically selects relevant modules and injects the content into the context provided to the model. By default, all added modules are eligible for selection, but users can control this by toggling modules on or off in the browser extension interface (\filledcircle{C1}). We render the activated modules at the bottom of each generated response as interactive chips (\filledcircle{D1}). Users can click on the chips to view the respective module's content. As shown in \filledcircle{D2}, users can remove chips that have been activated for a given message. Doing so is equivalent to toggling the module off using the extension's interface.

\section{Retrieving and Integrating Knowledge Modules}
\label{sec:router}
In this section, we detail Knoll's implementation. First, we describe how our router selects modules that are relevant to the user's query. Then, we discuss how we inject selected knowledge into the user's query directly on the commercial platform and in real-time. 

\subsection{Retrieving Relevant Knowledge Modules}
\label{sec:retrieval}
Since a user will have multiple modules enabled, determining which modules to use for a given query is critical. For example, if a user asks about new ideas in HCI, the system should include knowledge stored on \emph{interesting research trends} but not include information on \emph{travel}. A naive solution may be to include all the knowledge that the user has stored in every query, but this approach does not scale as the amount of knowledge that the user stores exceeds the size of the context window. To this end, we introduce a router that retrieves modules relevant to the query. Ablations on our router design are included in the Appendix~\ref{sec:app_tech}. 

Our module router employs a two-stage retrieval pipeline commonly employed in modern RAG systems~\cite{gao2023retrieval,glass2022re2g}. First, we retrieve a large pool of potentially relevant information from the activated modules; and then, we rerank the modules by relevancy, selecting those that are most related to the query. When designing our router, we prioritized \emph{recall} of knowledge, leveraging LLMs' capabilities to produce reasonable outputs even when given some amount of irrelevant information in the prompt~\cite{cuconasu2024power,wang2024astute}. Our retrieval pipeline is modeled after best practices, especially given the latency constraints we face in delivering a real-time system, but as always we assume that its components can be improved. With new advancements to retrieval-augmented LLMs, we expect that the pipeline used in Knoll will continue to improve.

\begin{figure*}
    \centering
    \includegraphics[width=1\linewidth]{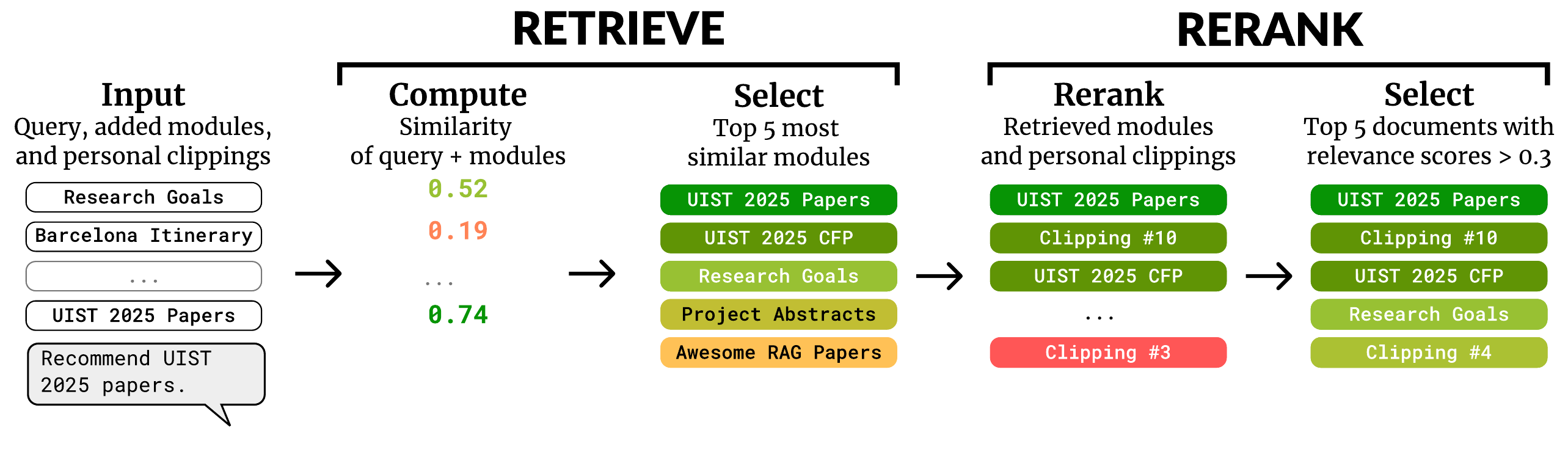}
    \Description[Diagram of Knoll’s two-step module routing process]{Diagram of Knoll’s two-step module routing process including retrieval and rerank steps. First, relevant modules and clippings are retrieved based on embedding similarity to the query. Second, documents are reranked using a relevance model, filtering those with scores below 0.3. The top documents are injected into the prompt sent to the model.}
    \caption{Our module router consists of a retrieve and rerank step. As input, we use the last two queries the user sent in the conversation, any added modules, and clippings in their Personal Module. First, we retrieve the top 5 most similar modules based on sentence embeddings. Then, we rerank the retrieved modules and clippings, filtering for the top 5 documents that have a relevance score $s > 0.3$. 
    }
    \label{fig:enter-label}
\end{figure*}

\subsubsection{Retrieving Relevant Knowledge}
Our first step is to retrieve knowledge relevant to the query from all of the information that the user has added to Knoll. We use the previous query that the user sent in the conversation, combined their current query, as input to our retrieval pipeline to retain relevant conversational context. 

We compute the embeddings of each module's content using Voyage AI's \texttt{voyage-3-lite} model~\cite{voyage2024embed}. We then calculate the cosine similarity between the module embeddings and the query embedding, selecting the five modules with the highest cosine similarities to the query. In this step, we handle the \emph{Personal Module} separately since the clippings are likely to include personal preferences that are frequently relevant to customizing the user interaction, and so, we retrieve all of the clippings without any filtering.

Concretely, we have a query $q$, embedding model $e$, a set of $n$ modules ($\mathcal{M} = \{m_1, \dots, m_n\}$), and a Personal Module $\mathcal{P} = \{p_1, \dots, p_t\}$ with $t$ clippings. Our set of retrieved documents $\mathcal{D}$ is formulated as follows where $k=5$: 
$$\mathcal{D} = \{m_i | m_i \in \mathcal{M}, \text{top-k similarity}(e(q),e(m_i))\} \cup \mathcal{P}$$

\subsubsection{Re-ranking and Filtering Documents}
The second step is then to re-rank the retrieved knowledge. First, we use a cross-encoder (Voyage AI's \texttt{rerank-lite-2})~\cite{voyage2024rerank} which, when given the query $q$ and our set of $n$ retrieved documents $\mathcal{D}$, predicts a relevance score, $s_i \in [0, 1]$ for each document $d_i$. We then select the top five most relevant documents, filtering out documents that are irrelevant (i.e., $s_i < 0.3$). Finally, we include any additional documents that are highly relevant (i.e., $ s_i \geq 0.7$) to the query. 

\subsubsection{Handling Large Modules}
Since modules do not have a predefined length, they may exceed the model's maximum context window. In this case, we will split the modules into smaller chunks, treating each chunk as a separate document $d_i$. To find where we should chunk the modules, we process all of the content into a Markdown format and recursively split the document into chunks based on user-defined headings (e.g., ``\#'', ``\#\#''), stopping once the chunk is able to fit in the context window. To retain information about where the chunk comes from, we prepend the the name of the document and the names of the higher-level headings. 

One limitation of our system is that the size of all activated modules must be less than 5 MB --- the default local storage threshold for Chrome browser extensions. This is equivalent to approximately $1.25\mathrm{e}{6}$ tokens. To manage the size of the Personal Module, we also set a storage limit such that each clipping must be under 500 KB in size or approximately $1.25\mathrm{e}{5}$ tokens in length. A private cloud-based solution would allow future versions of the system to scale beyond this limit.

\subsection{Injecting Knowledge into Queries}
Knoll incorporates the relevant retrieved knowledge into queries users send to an LLM in real-time. To integrate our system into users' workflows, we design Knoll such that the knowledge injection occurs directly on the commercial platform. 

\subsubsection{Integrating Knoll with Existing Platforms} 
Knoll runs directly inside existing LLM services (i.e., OpenAI's ChatGPT and Anthropic's Claude). When a user submits a query to the chatbot, our extension listens for the \texttt{fetch} request to the chat completion endpoint. We intercept that request and then call Knoll's module router to retrieve the relevant knowledge. We prepend the retrieved knowledge to the original prompt, overwriting the query in the original request. We then resume the \texttt{fetch} request, which proceeds to the model's server as if the user had directly entered the knowledge when submitting their query. When the server returns with the generated response, we post-process the reply to remove the injected knowledge so that it does not appear in the LLM's chat bubbles. Finally, the extension renders small user interface chips for each injected module as interface feedback to the user. See Appendix~\ref{sec:app_system} for more details.

\subsubsection{Running Knoll in Real-Time} 
Since our extension works in real-time, latency is an important consideration for system design. First, for our embedding (\texttt{voyage-3-lite}) and re-ranking model (\texttt{rerank-lite-2}), we selected smaller models that are designed for improved latency while still being performant. \new{To empirically validate this choice, we compare the recall and latency of different embedding models~\cite{reimers2019sentence, openaiembed}, and confirm that \texttt{voyage-3-lite} provides the highest recall score with a reasonable latency (see Appendix~\ref{sec:app_tech} for more details).} Second, we send a document $d_i$ only once per conversation as the model will retain the previously sent knowledge in its context. This decision reduces the amount of redundant information given to the model, which may hinder performance~\cite{wueasily}. However, with this approach, the model's performance may degrade as the conversation progresses and the context length increases, becoming more distanced from the initial knowledge injection. Finally, we cache module embeddings on the server-side. We stress-test our router, simulating 100 concurrent users and find that the mean latency is $281\text{ms} \pm 84$.

\subsection{Technical Implementation}
The Knoll Chrome browser extension and web application is implemented using React and Express.js. We use Firebase for user authentication and to store log data for our evaluation. Personal Modules are only stored on the browser extension's local storage. Otherwise, we fetch document contents using the Google Docs API and GitHub API each time the user issues a request to refresh their knowledge and store the text data in the browser extension's local storage as well. For integrating the knowledge modules into the user's prompt, our module router is implemented using a Flask backend server with a Celery task queue. The current version of Knoll integrates with ChatGPT and Claude, although the implementation can be extended to other web-hosted chatbots (e.g., Gemini, DeepSeek). 

\subsection{Boundaries and Errors}
Our architecture makes tradeoffs that can yield errors. One issue inherent with a community knowledge ecosystem is that users can import knowledge that contains knowledge conflicts within or between modules. For example, one module might advise avoiding a particular API or design pattern while coding, while another module might explicitly advocate for it. In our prompt, we ask that the model identify if there are any conflicts in the knowledge provided and share it directly with the end-users. However, we do not provide any measures to resolve knowledge conflicts---rectifying knowledge conflicts is an unsolved technical problem. Nonetheless, for Knoll, we chart two paths forward. First, for conflicts within modules, version control features built into our infrastructure, such as branching and merging, allow users to manage potential conflicts upstream. Second, finding technical methods for resolving these knowledge conflicts is an active area of research within NLP that may surface techniques which we can integrate into our infrastructure~\cite{wang2024astute,jin2024tug,wang2025retrieval}. As implemented, the language model will provide feedback to the end user when knowledge is in conflict (e.g., ``Some of the information that you provided to me suggests$\ldots$ but other information argues$\ldots$'').

Second, adding external knowledge can mislead models. Prior work found that while retrieval-augmented language models are significantly better at handling long-tail facts, their performance on standard tasks may degrade~\cite{mallen2023not}. One possible failure mode occurs when the model over-relies on the external information. In Knoll, this problem can manifest when users try to transition topics within a conversation; the model tries to apply modules that were previously provided as context but may no longer be relevant. 

Finally, as mentioned in Sec.~\ref{sec:retrieval}, we prioritize the \emph{recall} of knowledge modules. For example, in our router design, we set a more liberal threshold when selecting documents after the reranking step. This decision increases the likelihood that irrelevant information may be provided to the model. Although language models are capable of generating an appropriate response regardless, users may be confused when the user interface indicates that irrelevant modules were used to answer their query.

\section{Deployment and Evaluation}
\label{sec:field}
Evaluating Knoll requires us first to execute a \emph{technical evaluation} of the system's accuracy in retrieving and integrating relevant modules into queries. Beyond validating the technical capabilities of our system, the behavioral question we seek to answer is whether users actually make use of external knowledge modules and, if so, how this access changes the way that users interact with LLMs. To truly gauge Knoll's efficacy, we want to understand how users naturally engage with the system, and thus, we publicly deployed our research prototype. In the two months since launching Knoll, our system has been used by over 200 people for tasks ranging from managing emails about course logistics for a large introductory CS course to planning a vacation itinerary. Our deployment study was approved by Stanford University's Institutional Review Board.

\subsection{Public Deployment}
% Next, we discuss details related to the deployment of Knoll. 
This deployment of Knoll occurred in two phases. We first launched our system with students at our university before announcing Knoll to a more general audience. 

\subsubsection{User Recruitment}
\label{sec:recruitment}
We distributed Knoll to users via two approaches. First, we seeded modules for relevant topics at the university, which were advertised to users via listservs, public flyers, social media, and word-of-mouth. Second, we had a general launch, describing the system's features, that was shared via the authors' social media accounts on X/Twitter and Bluesky. We also posted on forums with active communities interested in LLM usage (r/promptengineering, r/ChatGPTPro, r/ChatGPT) and new technical products (HackerNews). To encourage naturalistic usage, there was no monetary incentive for joining Knoll.

\subsubsection{On-boarding} After downloading the extension, users consented to participate in our evaluation, were shown a brief tutorial on how to use Knoll, and then were given free rein to explore the system.\footnote{\url{https://knollapp.com/instructions}} They also had a ``Welcome to Knoll'' module automatically added to their extension, which they could use to ask questions about how to use the system. To maximize the ecological validity of our evaluation, we did not provide any instructions on which modules to download or mandate that users make new modules.

\subsubsection{Predefined Modules} To circumvent the cold-start problem, we seeded our ecosystem with an initial set of modules created by the authors. One set of modules involved local university knowledge, e.g., requirements for undergraduate majors and course reviews, allowing users to issue queries to their LLM such as ``Which courses should I take next term that fulfill requirements for the AI track in Stanford CS major?'' For the public deployment, we seeded modules with more general-facing knowledge (e.g., the upcoming CHI conference program, recent research trends), allowing users to issue queries such as ``Given my interests, which upcoming CHI sessions should I check out?''. As mentioned in Sec.~\ref{sec:recruitment}, we advertised a subset of these predefined modules to users.

\subsection{Collected Data}
From our consented participants in the public deployment, we collected the following data which we use in our evaluation: 
\begin{itemize}
\item \textbf{User Queries}: If at least one module has been activated, we stored the prompt that the user sent to ChatGPT / Claude as well as the names of the modules that were activated. To respect the privacy of users, we do not collect prompts for which modules were not activated, and thus unlikely to be relevant to our system evaluation. In addition, if knowledge from the Personal Module was retrieved, we only retain high-level information that this module was activated and do not log the exact clippings that were used. 
\item \textbf{Usage Logs}: To understand how users interact with Knoll, we also collect log data around clipping behavior (e.g., number of clippings added, number of clippings deleted) and sharing behavior (e.g., number of clippings shared). We also log how many modules the user has historically added as well as which modules are currently toggled on versus off. 
\item \textbf{Survey and Interview Data}: Finally, we collected survey data on users' experiences with Knoll (N=17). We conducted follow-up interviews with eight survey respondents.
\end{itemize}

\section{Technical Evaluation}
\label{sec:technicaleval}
Before understanding the impact that Knoll had on users, we begin with answering: does Knoll knowledge meaningfully change the responses from the LLM? Here, a highly performant system will be able to~(1) identify the modules related to the query and~(2) generate accurate responses that make use of relevant external information. We begin in this section with a technical evaluation of Knoll on a subset of user queries collected during our field deployment. \new{The goal of this evaluation is to validate that our approach can retrieve relevant modules and also integrate this knowledge to improve generated responses.} 

\subsection{Method}
From the 1,690 queries collected during the public deployment that had at least one module activated, we randomly sampled 100 as our test set. In total, the queries covered 16 different modules. For each query, we \remove{then generated a response using GPT-4o with the relevant module(s) inserted and a response using the baseline GPT-4o model.}\new{make two calls to the OpenAI API --- one with Knoll enabled and one without. When Knoll is enabled, we use the query as input to our module router and retrieve any relevant knowledge module(s); the query and modules are then sent to \texttt{GPT-4o}. For the baseline response when Knoll is not enabled, we send just the query to \texttt{GPT-4o} without any added information.}

We collected three human annotations per query. Judging the quality of the answers requires some background knowledge on the topic. Thus, we recruited two undergraduate students, two graduate students, and one of the paper's authors as annotators as many of the questions relate to either internal university topics. 

To assess the quality of the retrieval and generation process, we use the following metrics, which are commonly employed when evaluating RAG systems~\cite{es2024ragas,saad2024ares}:
\begin{itemize}
\item \textbf{Context Relevance:} Annotators label whether a returned module is relevant or not to the query (i.e.,``Select all modules that are relevant to the prompt.''). Across all three annotators, if at least one person labels a module as being related to the query, then we mark that module as relevant. This decision reflects our goal of maximizing recall: if even a single person finds the module useful, it suggests potential value to downstream users. We find annotators have a substantial agreement on relevance (Fleiss's $\kappa = 0.67$). We report the context relevance of the first module our pipeline returns (Relevance @ 1) and the relevance of returned documents (Relevance @ 5). 
\item \textbf{Answer Relevance:} Turning to generation quality, we measure how relevant the generated answer is to the query. An answer is considered relevant if it directly and appropriately addresses the original question. This measure focuses on how well the answer matches the intent of the question, without evaluating factual accuracy.
\item \textbf{Answer Quality (Win Rate):} We use preference as a proxy for answer quality. Annotators are shown the responses generated using a baseline \texttt{GPT-4o} versus with Knoll in a side-by-side manner and asked to select which response they prefer. We randomize the order that responses are displayed. Win Rate is the percentage of queries where annotators preferred responses generated with Knoll.
\end{itemize}

Finally, we select 50 queries for a more extensive annotation process used to benchmark \textbf{Recall}. For each query, annotators are presented with all 16 module options and asked to select \emph{all} modules relevant to the query. Similar to Context Relevance, we take the union of annotators' selections as our ground-truth. Annotators had moderate agreement on the task (Fleiss's $\kappa = 0.51$). Additional methodological details are in Appendix~\ref{sec:app_tech}. 
% \begin{table}[]
%     \centering
%     \begin{tabular}{lrr}
%     \toprule
%          &  Recall \\
%     \midrule
%     \texttt{GPT-4o} (Zero-shot) & $0.37$ \\ 
%     \texttt{GPT-4o} (Few-shot) & $0.21$ \\
%     \texttt{o3-mini} & $0.78$ \\
%     Ours & $0.76$ \\
%     \bottomrule
%     \end{tabular}
%     \caption{Caption}
%     \label{tab:my_label}
% \end{table}
\begin{figure}
    \centering
    \includegraphics[width=\linewidth]{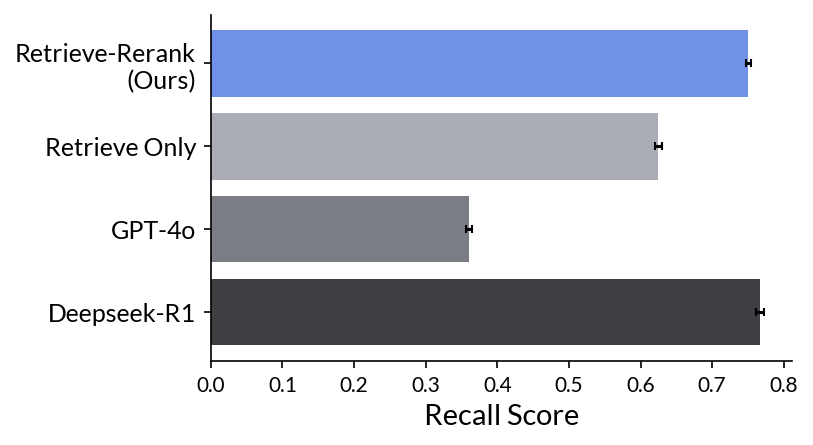}
    \Description[Horizontal bar chart showing Recall Score]{Horizontal bar chart showing the Recall Score for four different systems --- Retrieve-Rerank (Ours), Retrieve Only, GPT-4o, and Deepseek-R1. Retrieve-Rerank achieves a recall of approximately 0.75, outperforming Retrieve Only (recall of 0.62) and GPT-4o (recall of 0.36). Retrieve-Rerank performs in-line with Deepseek-R1 (recall score of 0.77).}
    \caption{Our two step retrieve-rerank router is able to recall the relevant modules for users' queries. It outperforms the ablated router with only the retrieval step (Retrieve) and directly using GPT-4o to classify if modules are relevant; the approach performs in-line with using a state-of-the-art reasoning model (Deepseek-R1~\cite{guo2025deepseek}) to classify if modules are relevant. $95\%$ CIs are computed by bootstrapping over 1,000 iterations.}
    \label{fig:recall}
\end{figure}
\renewcommand{\arraystretch}{1}
\begin{table}[b]
\small
    \centering
    \begin{tabular}{lrrr}
    \toprule
    &  \textbf{\makecell{Target\\ Value}} & \textbf{\makecell{Technical\\ Eval.}} & \textbf{\makecell{User\\ Eval.}} \\
\midrule
   % \emph{Generation} \\
   % \cmidrule(rl){2-5} 
   Win Rate (Relevant Info.) & $100.0\%$ & $81.5\% \pm 6.4$ & $81.5\% \pm 14.7$\\
   Win Rate (Distracting Info.) & $50.0\%$ & $54.8\% \pm 7.9$ & $52.0\% \pm 19.6$\\ 
   % \rule{0pt}{3ex}  
   % Retrieval\\
   % \cmidrule(rl){2-5} 
   Context Relevance@1 & $100.0\%$ &$49.0\%\pm9.9$ & $64.3\%\pm15.5$\\
   Context Relevance@5 & $100.0\%$ & $43.0\%\pm8.3$ & $63.3\%\pm 13.4$\\
   \bottomrule
    \end{tabular}
    \Description[Evaluation results table comparing Knoll and GPT-4o]{Evaluation results comparing Knoll with baseline GPT-4o on metrics such as win rate and context relevance. Knoll outperforms the baseline in win rate when relevant knowledge is available (81.5\%) and performs similarly when knowledge is distracting.}
    \caption{Knoll improves answer quality when external knowledge is relevant and performs comparably to \texttt{GPT-4o} when irrelevant or``distracting''~\cite{cuconasu2024power} modules are included.  We report win rate on the subset of queries where modules are relevant, win rate on the subset of queries where modules are unrelated, and context relevance. Technical evaluation comes from annotator labels; user evaluation come from survey responses. $95\%$ CIs are computed via the Wald interval (Win Rate) and bootstrapping over 1{,}000 iterations (Context Relevance).}
    \label{tab:technical}
\end{table}

\subsection{Results}
\subsubsection{Knoll retrieves relevant knowledge modules} \new{We start by validating that our approach can retrieve all of the information relevant to the query. To do so, we compare the recall score of our proposed approach with alternative module router designs. The full retrieve-and-rerank router that we use in Knoll obtains a recall score of $0.75 \pm  0.05$. As shown in Fig.~\ref{fig:recall}, this design outperforms using an ablated router with only the retrieval step ($0.63 \pm 0.08$), and using an LLM (\texttt{GPT-4o}) to classify whether a module is relevant to a given query ($0.36 \pm 0.07$). A two-sample $t$-test confirms that our approach has a significantly greater recall score compared to the retrieval only design ($t=40.7$, $p < 0.001$) and the \texttt{GPT-4o} classifier ($t=137.4$, $p < 0.001$). In addition, our router performs in line with a state-of-the-art reasoning model, \texttt{DeepSeek-R1}~\cite{guo2025deepseek}, which obtains a recall score of $0.77 \pm 0.09$. Although 
using a reasoning model offers better recall ($t=-5.09$, $p < 0.001$), it does come at the cost of higher latency compared to our method.}

\subsubsection{Knoll produces more relevant and higher quality answers} 
\new{So far, our evaluation has focused on validating that Knoll can recall relevant modules, without considering how including this external knowledge impacts the generated responses. To answer this question,} we compare the relevancy and quality of answers generated with Knoll enabled versus that of \texttt{GPT-4o}. \new{In this setup, to outperform the baseline, Knoll must not only retrieve the relevant knowledge modules but also ensure the content is presented in a way such that the model can make use of the information when generating responses. In cases where we retrieve irrelevant modules or the model ignores the inserted information (e.g., providing too much context leading to the ``lost-in-the-middle'' phenomenon~\cite{liu2024lost}), Knoll's answer relevance and quality will suffer compared to that of the baseline.}

To start, we find that most answers tend to be relevant to the query. However, answers generated using Knoll are consistently rated as more relevant ($90.0\% \pm 3.4$ of answers) compared to those generated by the baseline ($71.7\% \pm 5.1$ of answers). A two-proportion $z$-test confirms that answer relevancy with Knoll enabled is significantly greater than that of the baseline ($z=5.70, p < 0.001$). Here, answer relevancy captures only whether the response addresses the question and does not consider whether it contains correct or useful information, leading us to also consider whether users prefer responses generated with Knoll.

Next, we analyze the quality of generated answers. As shown in Table~\ref{tab:technical}, our win rate is $81.5\%\pm 6.4$, indicating that annotators prefer Knoll's responses over the baseline for $81.5\%$ of the queries that require external knowledge. We conduct a one-proportion $z$-test to confirm that our win rate is in fact significantly higher than random chance ($z=9.51$, $p<0.001$). \new{This result indicates that models are able to incorporate the external information that has been retrieved, thus improving the quality of the generated responses.}

\subsubsection{Knoll performs on-par with baseline even with irrelevant context}
\label{sec:recall}
When users interact with LLMs, many queries may not require external knowledge from the modules. It is therefore important to evaluate performance on queries where external knowledge may be irrelevant. \remove{As discussed in Sec.~\ref{sec:knoll}, we prioritized module recall when designing Knoll, and we achieve a recall score of $0.75$. This outperforms using \texttt{GPT-4o} to classify whether the module is relevant to the query ($0.37$) and in-line with a state-of-the-art reasoning model (\texttt{DeepSeek-R1}~\cite{guo2025deepseek}) which obtains a recall score of $0.77$.} \new{As discussed in Sec.~\ref{sec:retrieval}, we prioritized recall when designing Knoll to ensure that all modules relevant to the query are included.} However, prioritizing recall also increases the likelihood of injecting irrelevant information. As shown in Table~\ref{tab:technical}, we observe a Context Relevance@1 of $49.0\%\pm9.9$ and Context Relevance@5 of $43.0\%\pm8.3$.

% \sout{As discussed in Sec.~\ref{sec:knoll}, we prioritized module recall when designing Knoll, and we achieve a recall score of $0.76$. This outperforms using \texttt{GPT-4o} to classify whether the module is relevant to the query ($0.37$) and in-line with a state-of-the-art reasoning model (\texttt{DeepSeek-R1}~\cite{guo2025deepseek}) which obtains a recall score of $0.77$.} 

But does including irrelevant context impact the response quality? To answer this, we consider the win rate for queries in our evaluation set where the external knowledge is \emph{not} relevant. We find that the win rate is approximately random, with annotators selecting the response from Knoll $54.8\% \pm 7.9$ of the time. This suggests that even when distracting information is injected into the prompt, the model is still able to produce high quality answers that are indistinguishable from responses generated when no irrelevant context is added. These results indicate that overall, Knoll is able to incorporate relevant external knowledge and produce better responses without harming answer quality on queries that do not require external knowledge. Nonetheless, errors may still occur; as such, we provide qualitative examples of successful and error use cases in Table~\ref{tab:examples}.

\subsubsection{External knowledge is not always useful, even when relevant to the query} 
We examined cases where adding relevant external knowledge did not improve response quality. For some queries, adding external knowledge related to the query actually distracted the model, leading to less relevant answers. For example, when a user asked for summer research internships through a university program, Knoll added a module with application tips. As a result, the response focused on application advice rather than internship recommendations. Another limitation arises when users ask about information not covered in any module. Doing so leads the model to ``fail loudly'' or reference its lack of relevant knowledge, potentially confusing users. Response quality is also influenced by factors beyond Knoll’s technical performance. For instance, when a user requested help drafting an internship application, the system correctly activated the module with application tips and incorporated example content. Yet, all three annotators still preferred the baseline response. In this case, even though the technical components of the system worked as intended, the lower quality of the module wound up undercutting the response. 

\renewcommand{\arraystretch}{1} % Default value: 1
\begin{table*}[]
    \centering
    \small
    \begin{tabular}{l>{\raggedright\arraybackslash}p{1.4in}>{\raggedright\arraybackslash}p{1.35in}>{\raggedright\arraybackslash}p{1.35in}>{\raggedright\arraybackslash}p{1.4in}}
    \toprule 
         &  \normalsize \textbf{Successful Cases} & & \normalsize \textbf{Error Cases} &\\
    \midrule 
    \normalsize Query &  ``Generate practice questions related to the course concepts on tools and toolkits.'' &``Advice on when / how to schedule quals '' & ``Is CS 123 offered in the spring?'' & ``I would like to know more about papers relevant to personal information management.''\\
    \midrule
     \normalsize Response & \emph{Here are exam questions tailored to terminology from your course: \newline\newline
    1. What is the main goal of prototyping tools? \newline
    Explain with an example how a prototyping tool might support rapid iteration in the design process \dots} & \emph{In CS, the qual is typically a depth exam in your research area, usually an oral exam with a faculty committee. Generally 3–4 faculty members and at least one member must be from the CS department. Set a date at lease one month in advance\dots} & \emph{Based on the knowledge provided, I don't see CS 123 listed among the CS courses offered. The knowledge contains a comprehensive list of CS courses, but CS 123 is not mentioned in this list.} & \emph{Great question! Based on the CHI 2025 papers provided, several papers are directly relevant to personal information management \dots}\\
    & & & \textcolor{Periwinkle} {\textbf{Information about CS 123 is not in the module, leading the model to ``fail loudly,'' explicitly mentioning the lack of provided information in the response.}} & \textcolor{Periwinkle} {\textbf{The user asks about relevant papers more generally, but the model only draws from the provided CHI 2025 Papers module when generating a response.}} \\
    \bottomrule 
    \end{tabular}
    \Description[Table of Knoll usage examples]{Examples of Knoll usage, showing four queries with summaries of successful and error responses. Successful cases include course-specific practice question generation and personalized paper recommendations. Error cases arise when the system lacks relevant knowledge or over-relies on provided content.}
    \caption{Sample queries illustrating successful use cases of Knoll (left) and error cases (right) in which the module lacks information on the query or the model over-relies on the provided external knowledge. \textcolor{Periwinkle}{\textbf{Bolded purple text}} at the bottom of the response indicates rationale for why the errors occur.}
    \label{tab:examples}
\end{table*}

\begin{figure*}[t]
    \centering
    \includegraphics[width=\linewidth]{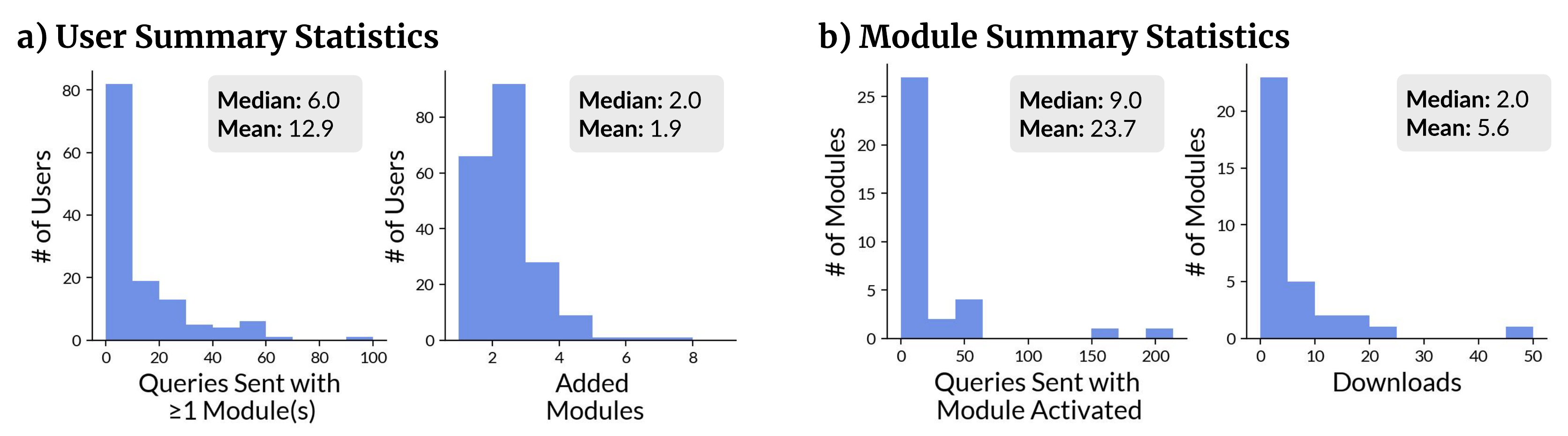}
    \Description[Bar chart showing number of queries per user]{Bar chart showing the number of queries sent per user with more than one module activated. Median is 6.0 and mean is 12.9, indicating skewed distribution with several high-use users.}
    \caption{We visualize summary statistics on a) our 203 Knoll users and b) the 39 modules that users added. For users, we show the number of queries sent with more than one module activated and number of modules added to the extension. For modules, we show the distribution of queries sent with a module activated and the number of times a module was downloaded (i.e., added by a user). We excluded the ``Welcome to Knoll'' module when reporting the number of downloads per module as it came pre-installed.}
    \label{fig:metadata}
\end{figure*}
\begin{figure*}
    \centering
    \includegraphics[width=\linewidth]{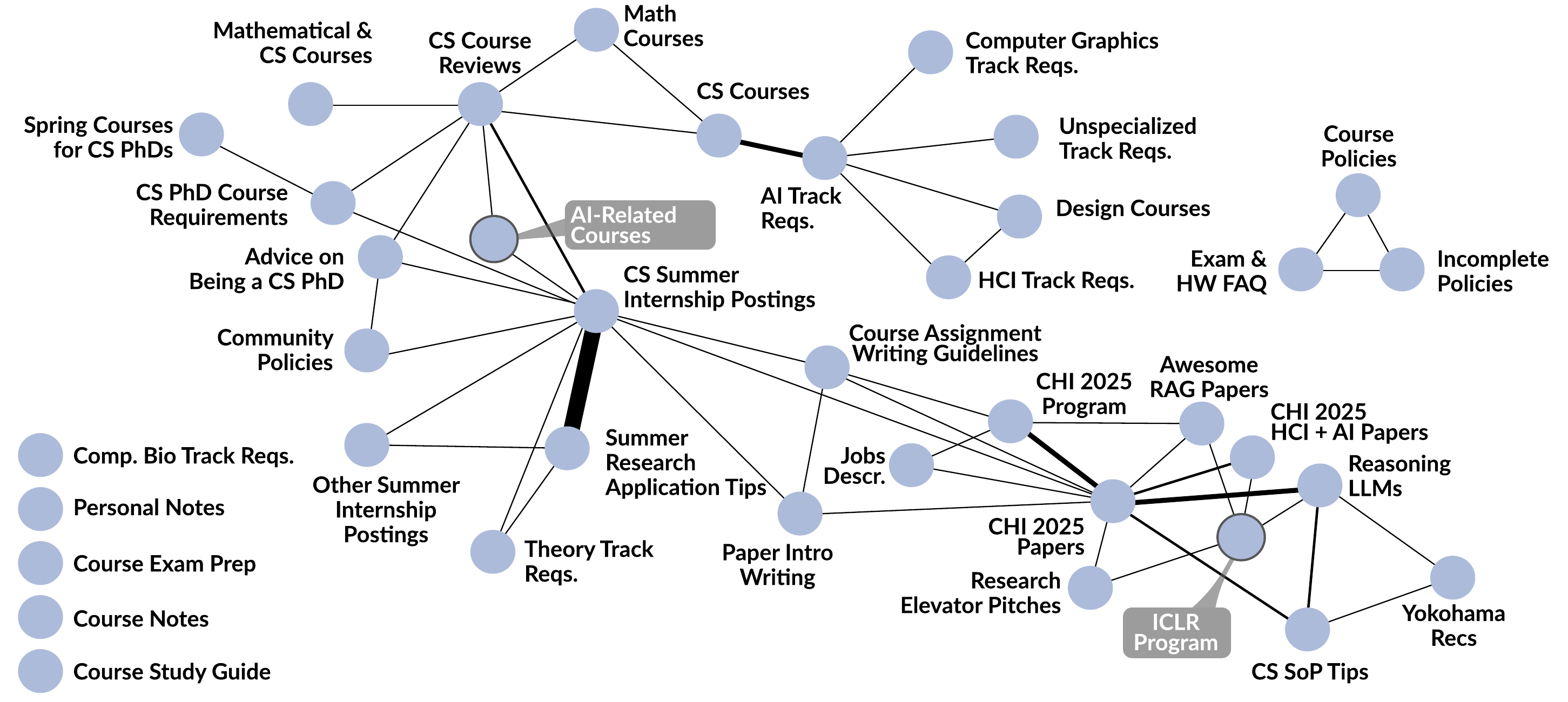}
    \Description[Network visualization of module relationships]{Network visualization of module relationships. Nodes represent different knowledge modules (like "CS Course Reviews", "CHI 2025 Papers", etc.) and edges indicate when users have both modules installed, with edge weight corresponding to the number of users with both modules. Most nodes are connected, with the strongest connection between the modules "CS Summer Internship Postings" and "Summer Research Application Tips."}
    \caption{We visualize which modules users have installed. In our graph, each node represents a module. Each edge indicates that a user has both modules installed with the edge weight corresponding to the number of users that have both modules installed. We exclude the ``Welcome to Knoll'' module from this analysis.}
    \label{fig:graph}
\end{figure*}

\section{User Evaluation}
Next, we explore how users interacted with the system during the public deployment. We analyze how users engaged both as knowledge consumers and creators as well as how having Knoll available shaped users' experience with LLMs.

\subsection{Method}
To better understand how users interact with Knoll, we triangulate between three sources of data: logs from all 203 users, qualitative survey responses (N=17), and semi-structured interviews (N=8). 

\subsubsection{User Recruitment} To recruit participants for our survey and interviews, we rendered a banner at the top of the ChatGPT / Claude interface, inviting our users for an optional survey after they had the extension installed for over one week. This process yielded N=22 users who both signed up and sent at least one message with a Knoll module activated. In total, 17 users completed the survey; we then conducted a follow-up interview with eight survey respondents. Participants were compensated with a \$10 Tremendous gift card for the survey and an additional \$20 Tremendous gift card for the interview.

\subsubsection{Survey and Interview Details} In our survey, we ask the users to provide preference data between a response generated using Knoll versus a baseline \texttt{GPT-4o} model for a subset of queries that the participant had sent (Win Rate) and labels on whether an activated module is relevant to the query (Context Relevance). They also provided short-answer descriptions on the types of queries they sent while using Knoll and opinions on the best / worst parts of the system. Finally, participants listed any other types of external knowledge they would want an LLM to have access to. 

From our interviews, we sought to understand why users sent certain queries to Knoll and their experience with the system. We showed participants queries they sent with Knoll enabled, asking their intention behind sending the query as well as their desired model response. We concluded by discussing their experience creating knowledge modules and overall reflections on Knoll. 

\subsection{Results}
\subsubsection{Users made use of a wide range of knowledge modules}
In total, 203 users downloaded and created an account with Knoll. Of these users, 135 submitted at least one query to ChatGPT / Claude with the extension installed. Overall, $14{,}378$ queries were sent over the two-month period with a median of $17.0$ queries per active user. $11.8\%$, or $1{,}690$ queries, had at least one module activated. 

The number of modules added ranged across users. As shown in Fig.~\ref{fig:metadata}, the median number of added modules is $2.0$. This suggests that most users downloaded Knoll to use a specific knowledge module (likely one that we advertised to them). However, there were also other users that added a wider range of modules to their extension with one user adding seven modules to their extension. This usage pattern illustrates how modules can help support a diverse range of queries that a user may send to an LLM. For example, one user imported five modules, which covered upcoming conference papers, how to write reading responses for their course, and open summer internship positions. We visualize the network of modules that users installed in Figure~\ref{fig:graph}.

Finally, in addition to modules that we promoted in our advertisements to users, we observed that a subset of 22 users made use of other module offerings listed in the public gallery on our Explore Page. The most popular of these modules contained content on course reviews for a university's computer science department and another covered recent trends in reasoning LLMs. Detailed information about modules are in Appendix~\ref{sec:app_user_eval} Table~\ref{tab:modules}. Via exploration, users can discover useful modules that they would not have thought of on their own. As P15 reflected, ``\emph{I think it's nice with the modules created, like you provide an a solution for something I didn't really know I needed.''}

\subsubsection{Users engaged in creating knowledge modules}
Users actively engaged as knowledge creators, contributing new modules through both document imports and content clipping. In total, they created nine modules, which included personal materials (e.g., notes, job postings, study guides) and content aimed at a broader audience (e.g., course policies, advice on writing academic paper introductions). \new{These newly created modules ranged in size from 1,500 to 11,000 words}. For instance, P3 described creating a module to support writing paper introductions: ``\emph{I am trying to gather the knowledge that I would gather myself... If I can log it into a module and send it to an LLM, the model would know what I know—otherwise the response would be too general}.'' Log data also showed that 11 users added to their Personal Module through clipping or manual entry. While most modules were for personal use, some sharing behavior emerged, with four users using the clipping-sharing feature.

\begin{figure}[tb]
    \centering
    \includegraphics[width=\linewidth]{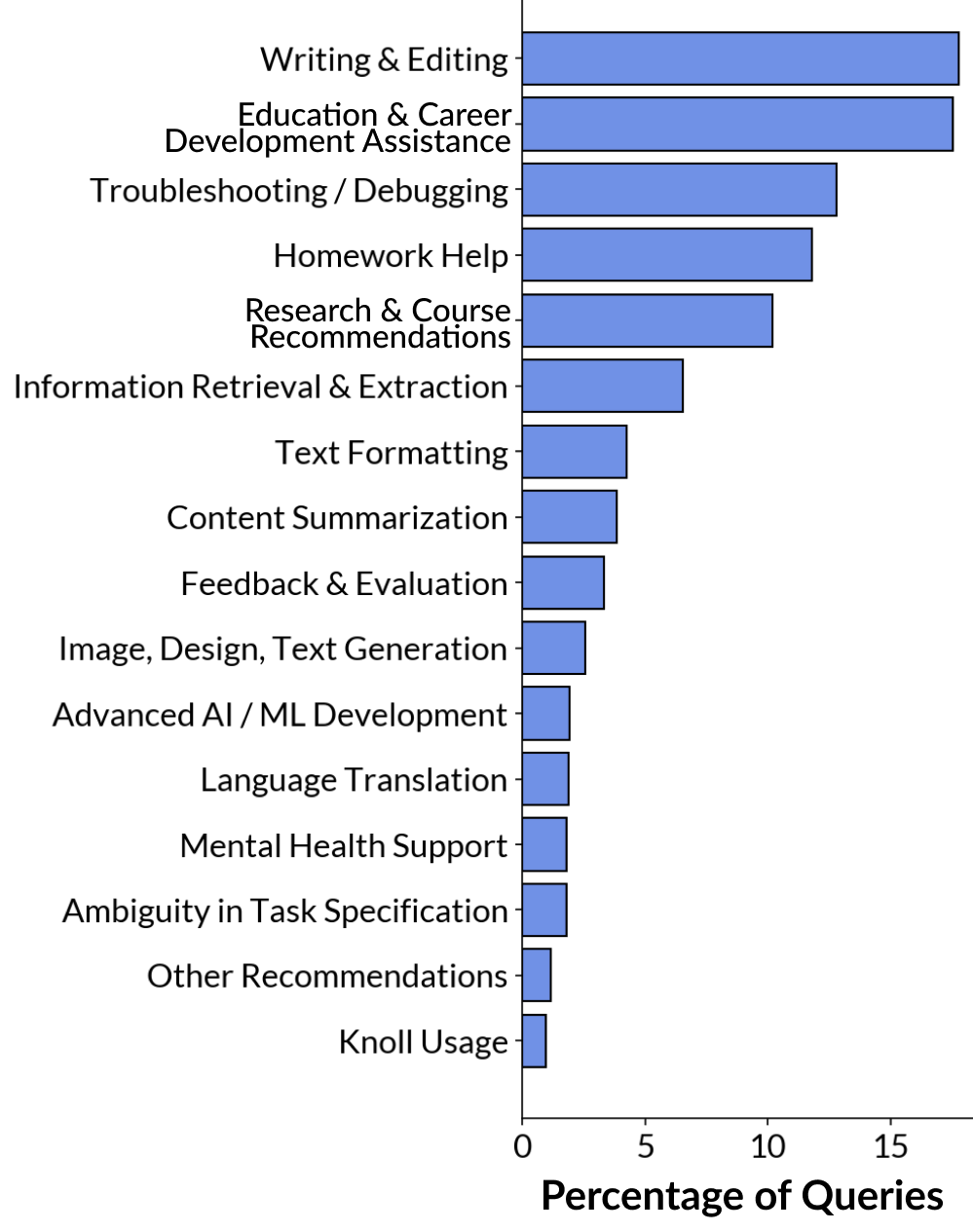}
    \Description[Bar chart showing query type distribution]{Bar chart showing distribution of query types. Bars represent percentage of queries, with categories like "Writing & Editing" and "Education & Career Development Assistance" having the highest percentages, followed by other categories including "Troubleshooting/Debugging", "Homework Help", and "Research & Course Recommendations".}

    \caption{Users employ knowledge modules for a diversity of tasks, including for tailoring their writing, receiving personalized recommendations, and seeking advice. We visualize the high-level task categories for all queries sent with at least one module activated.}
    \label{fig:message_types}
\end{figure}
\subsubsection{Knoll supports a diversity of query types}
\label{sec:cluster}
Next, we explore what types of queries users sent when using Knoll. Retrieval systems are often benchmarked on question-answering tasks~\cite{xiong2024benchmarking, chen2024benchmarking}, but in our naturalistic usage setting, we expect that modules are useful for a wider range of tasks, including personalized recommendations~\cite{wu2024coral} and text generation~\cite{li2022survey,cheng2023lift}. To answer this question, we analyze the queries that users sent that had at least one module activated. Following prior work~\cite{tamkin2024clio}, we first extract the task that the model is being asked to perform for each query. Then, we obtain the semantic embedding of the extracted task,  using $k$-means clustering where $k$ is chosen such that the average number of queries per cluster is $40$, allowing us to fit all of the queries in the context window of the model during the summarization step; this results in a total of $39$ clusters. For each cluster, we generate a title and short two-sentence description using the extracted tasks as input to an LLM. Finally, we manually de-duplicate clusters, resulting in 16 higher-level groups. See Appendix~\ref{sec:app_query} for more methodological details.

We find that modules are used to provide context for generative tasks including writing and editing, advice-seeking, and recommendations spanning different topic areas. To illustrate the types of queries users send to Knoll, we use the ``Educational \& Career Development'' cluster as an example.\footnote{To preserve user privacy, we provide paraphrased versions of the queries submitted to Knoll.}  In this cluster, we observed conversations asking for advice about applying to internships (``\texttt{What relevant experiences should I highlight in my application to this summer research project on AI agents}''), receiving recommendations for course selection (``\texttt{What are the easiest courses I can take to fulfill the AI track?}''). While most of the queries were related to text generation, there were also prompts asking for factual recall over knowledge modules, such as ``\texttt{Is CS 101 offered in the spring?}'' or ``\texttt{Who is the contact for this summer research internship?}''.   

\subsubsection{Knoll enables more customized interactions}
Knoll addresses a data void problem~\cite{boyd2018data}. Users reported that with Knoll they asked LLMs to perform tasks they otherwise would not have. For example, P3 used Knoll to get personalized recommendations on conference papers --- a task they would not have done prior to using the system: ``\emph{I don't think the LLM will know about [the conference] since it is really recent...and even if the model does know, it's very likely it will hallucinate.}'' As P4 reflected, ``\emph{I could ask it questions, assuming context that ChatGPT or Claude would not know, and this was more convenient.}'' Overall, using Knoll improved the quality of model responses. As shown in Table~\ref{tab:technical}, participants preferred responses generated by Knoll in $81.5\%\pm14.7$ of cases.

Users also noted that Knoll simplified the process of providing relevant information to an LLM. Previously, some users avoided certain queries, while others manually added the necessary context into their prompt or uploaded files. With Knoll, P9 commented that they ``\emph{think it's good that you can maintain a document as a database so you don't need to manually input it into ChatGPT.}'' P5 also noted that using Knoll reduced a previously laborious process: ``\emph{if it's a task I need to do every day or like I'll keep repeating similar tasks, it is worth building such modules.}''

\subsubsection{Knoll complements existing model features} \new{When using Knoll, users still had access to features available on the commercial platform, such as LLM-enabled search and file uploads. Having access to localized knowledge via Knoll complemented or even augmented how these features are used. For example, several users uploaded their resumes and transcripts when using modules that helped with applying to internships or finding courses. As P11 explained, while they used ChatGPT search extensively, they viewed Knoll as an opportunity to ``\emph{search through information that isn't necessarily Google searchable but still things that are community based}.'' Although P11 planned to continue using LLM-enabled search for topics available on the web, Knoll made different types of knowledge accessible to them, opening up the possibility for new forms of interaction. } 

\subsubsection{Users expressed concern about hallucination and module content quality}
One concern users had when using Knoll was with the quality of the module content. In order for users to trust responses generated with Knoll, they must first have confidence in the modules' content. While users can view the source content, it is unlikely that users will read through all the modules they want to add, especially given that some are upwards of 150{,}000 words. We observed that some users reported feeling uncertain about the quality of modules, particularly when the modules were authored by other users. As P16 explained, ``\emph{because I was not the one who created the module, I wasn't sure what information it had, so it wasn't clear to me what the model actually knew.}'' P16 further reported that when the model provided a response that seemed wrong to them, they were unsure if the model was hallucinating or if there was information in the imported module that was incorrect. 

Knoll is not a panacea for model hallucination. In fact, users are likely to be less tolerant of hallucinations with Knoll enabled. As P15 stated, ``\emph{maybe that's unfair of me, but I guess for me the whole purpose of using this extension is to reduce hallucinations and draw upon very specific knowledge... then there would be no additional benefit of using Knoll if it hallucinates similarly to when I use it without the extension.}'' When using off-the-shelf models, users have more tolerance for hallucinations as they assume the model may not have access to specific knowledge, but with Knoll enabled, users have a higher expectation for the model's performance.

\section{Discussion}
\label{sec:discussion}
In this section, we detail additional applications of modules and design considerations that arise when creating a knowledge ecosystem for language models. We conclude with a discussion of limitations and future work. 

\subsection{Extending the Knowledge Ecosystem}
Through our field deployment, we saw users engage with Knoll to support different tasks mainly focused in academic and research-oriented settings. However, knowledge modules have vast applications that expand beyond those surfaced in our initial deployment. One application that several survey respondents were interested in was incorporating more socially sourced knowledge such as interest groups. For example, both P11 and P17 mentioned that much of the knowledge they have about their school comes from a university-specific social media app or from talking with their friends --- both of which are inaccessible to language models. P17 imagined having ``\emph{a tips database that is being updated by all the students somehow with whatever people know, they can just like maybe write it there and contribute}'' or a module that could be up-to-date with social events near them. 

Another application area is for incorporating information from communities that may be underrepresented on the Internet and less likely to be represented in the training data. For example, prior work has shown that LLMs struggle with everyday cultural knowledge that comes from non-Western regions (e.g., ``what is a common snack at sports stadiums in South Korea?'')~\cite{myung2024blend}. Ma et al.~\cite{ma2024evaluating} also found that LLMs used for providing mental health support fall short for LGBTQ+ users, who find the responses to be overly generalized and unable to adequately handle identity-related questions. In these scenarios, we can imagine users, who have relevant cultural knowledge or experiences, creating knowledge modules to share with others in their community. Moreover, allowing users to author their own knowledge modules allows them to augment models with this information while retaining control over the information being added and how their community is being represented~\cite{smart2024socially}.

\subsection{Design Considerations}
\subsubsection{Module Content}
One of the key design considerations for Knoll is how we manage module content. Two aspects that we discuss are quality and granularity. First, as several survey respondents reported, the quality of module content influences how much they trust the responses generated and shapes how they interact with the system. For example, P6 reported that they only selected modules where they thought they had enough expertise to validate whether the generated responses were accurate. We also observed respondents employ their own assurance checks over module quality, such as skimming the source content before installing or submitting test queries to first ascertain whether the module contained the knowledge they wanted. To make this process easier for users, future designs could incorporate verified modules that come from trusted sources, sample responses that showcase what types of queries the module supports, or a rating and review system. 

Second, we do not impose any restrictions on the granularity of module content. As a result, there are many modules that cover closely related topics (e.g., there were modules on papers at CHI 2025, AI-related papers at CHI 2025, and the CHI 2025 program). This design choice allows us to capture a plurality of viewpoints on a topic. However, the trade-off is that it may be cumbersome for users who have to search for the modules they want in a sea of similar options.

\subsubsection{Data Ownership and Access}
Another important consideration is with regards to data ownership. As P3 remarked, when creating their module they were worried ``\emph{if I am doing something wrong because it doesn't feel right to like just copy paste other people's stuff.}'' While P3 ultimately decided it was permissible since the data they were accessing was already publicly available and they personally knew the data creators, their concerns are indicative of a larger question around data usage. The benefit of Knoll is that the module creation process is user-driven rather than being top-down and extractive, as is the status quo with most NLP datasets~\cite{bird2024must}. Users have more ownership when it comes to adding their own data. However, questions of consent and copyright are still germane when it comes to incorporating data created by others into their modules. 

\subsubsection{Module Governance}
As our ecosystem grows, governance and safety become a critical issue. What happens if modules contain malicious or harmful data? While the guardrails that commercial models have in place may stymie some concerns, prior work has shown users can still circumvent these safety measures~\cite{qi2024fine}. There is also the possibility of less apparent but still pernicious attacks in the form of ``adware'' or phishing attempts that can be included in modules~\cite{tao2023opening}. For example, a user might install a Knoll module, only to find that their LLM starts endorsing a particular brand of shoes in its responses. Although users can view the source content of modules, this vetting process is tedious. Overall, there are trade-offs between having a more decentralized ecosystem that affords end users more autonomy in determining what knowledge they can add to an LLM versus having a more centralized body that determines whether modules fit a set of community standards. 

\subsection{Limitations and Future Work}
Finally, we address limitations of our system and provide suggestions for how future work can expand on the infrastructure introduced in this paper. 

\subsubsection{Supporting other forms of knowledge}
First, in our field deployment, many Knoll users gravitated towards educational or research-oriented use cases --- likely a result of our recruitment strategies. However, this represents only a sliver of the breadth of tasks that knowledge modules can be applied to. Future work can explore how Knoll generalizes, extending modules to cover other domains and interest areas. \new{For example, conducting targeted deployments of Knoll in specific communities can yield insights into the different types of localized knowledge (and thus what modules) people want to make use of and how this then shapes their interactions with models.}

Second, our system focuses on knowledge that can be provided as text input. While Knoll accommodates some unstructured forms of data, it does require that the knowledge be explicitly written down. We recognize that this excludes forms of knowledge that are tacit or more difficult to verbalize~\cite{nonaka2009knowledge}. Future work can explore methods for externalizing tacit knowledge --- or the process of transforming this unspoken knowledge into an explicit, documented form~\cite{nonaka2009knowledge} --- and creating modules from this articulated information.

\subsubsection{Exploring new interactions with Knoll} \new{
This work only begins to explore the new forms of interactions enabled by our modular knowledge ecosystem. Knoll provides new interaction surfaces for users to curate external knowledge their language model can access, opening up new tasks by addressing data voids. We imagine that, beyond this, there are new interactions for both creating and importing modules. On the knowledge creation side, our system allows users to either manually update or clip knowledge. Future work can explore how we can automatically infer knowledge updates via conversation or have proactive agents that can curate candidate modules for the user. For knowledge consumers, our system is designed to treat all selected modules as equally important; however, we examine new ways in which users can communicate how modules are prioritized or combined.
}

\subsubsection{Data privacy with commercial models} While Knoll allows users to control who has access to their knowledge modules, all modules are still sent to third-party services (ChatGPT, Claude) when knowledge is added to the user's prompt. Users may not feel comfortable creating modules that contain private information, which they do not want to expose to commercial providers. However, relying on third-party models is not required. Our infrastructure can be expanded to included local language models that can even be hosted on the user's machine, allowing them to create modules with more private information. 

\subsubsection{Technical limitations} 
There are technical limitations with our current implementation of Knoll. First, since Knoll directly incorporates knowledge onto the ChatGPT / Claude interface, we face platform limits in the amount of context that can be added. These limits also differ depending on what subscription the user has (e.g., ChatGPT Pro allows more content to be added into the message than the Free tier allows). In addition, since we only inject knowledge the first time a module is triggered in a conversation, the model may ``forget'' the knowledge as the context window length increases~\cite{laban2025llms}. Users can start a new conversation in which the knowledge will be added again, but this interaction may be cumbersome. An alternative design choice here could have been to inject relevant knowledge for each message in a conversation; however, this design may dramatically increase the context window length. As language models become more adept at handling long contexts, we expect this to be less of a concern. Finally, knowledge modules only support text at the moment. Given the improving multi-modal capabilities of models, we can imagine supporting modules of different modalities or using models to convert sources into a text format.   

\section{Conclusion}
In this work, we propose creating an ecosystem of knowledge modules that allows users to augment LLMs with localized information. We build Knoll, a software infrastructure, to support this vision, enabling end users to easily customize language models using these modules. We publicly deploy Knoll, reaching over 200 users in two months for a diverse range of tasks. Our evaluation shows that users actively engage in the knowledge ecosystem, employing Knoll to improve the quality of model responses and contributing new knowledge modules. As our ecosystem grows, we hope not only to encompass a wider breadth of localized knowledge but also to foster a community where end-users can build and share knowledge together.  

%TC:ignore
\begin{acks}
We thank Jordan Troutman, Yutong Zhang, Andrew Wang, Charlotte Zhu, Omar Shaikh, Michelle Lam, Caleb Ziems, Michael Ryan, Lindsay Popowski, and Helena Vasconcelos for their helpful feedback on this work. We also thank members of the Stanford HCI group and SALT Lab for helping beta-testing our extension. This work was supported in part by IBM as a founding
member of the Stanford Institute for Human-centered Artificial Intelligence (HAI), a Stanford HAI Seed Grant, the Sloan Foundation, and NSF grant IIS-2247357. Dora Zhao is supported in part by the Brown Institute for Media Innovation.
\end{acks}

\bibliographystyle{ACM-Reference-Format}
\bibliography{ref}
\appendix
\section{System Architecture Details}
\label{sec:app_system}
We prepend the following prompt to the user's query along with the retrieved content from the knowledge modules. We employ Chain-of-Thought prompting techniques~\cite{wei2022chain} and instruct the model to identify knowledge conflicts as well as ignore irrelevant information:

\aptLtoX[graphic=no,type=html]{}{\begin{quote}}
\begin{verbatim}
 You are a helpful and knowledgeable assistant
 that provides answers to a user's query.

 We provide additional knowledge that might
 be helpful for answering the query.

 Let's think step by step. 

 1. Check whether the knowledge is relevant to
 the query. If the knowledge is relevant,
 incorporate it when answering.

 2. If the knowledge is NOT relevant, disregard
 the knowledge, do NOT make reference to it,
 and answer the query. Ignore information that
 is irrelevant. Do NOT search the web if you
 have sufficient knowledge.

 3. Check whether there are conflicts in
 the knowledge. Report conflicts in the output
 if they exist.

 Knowledge: ${MODULE CONTENTS}$
\end{verbatim}
\aptLtoX[graphic=no,type=html]{}{\end{quote}}

\section{Additional Technical Evaluation}
\label{sec:app_tech}
We evaluate different versions of our retrieval pipeline to validate our design decisions. 

\paragraph{Benchmarking Recall} We evaluate the \textbf{Recall} of our pipeline. To do so, we run our pipeline with the 16 modules mentioned in Sec.~\ref{sec:technicaleval} as input. Our pipeline only returns five modules per query, so we cap the number of relevant modules at five when computing recall. Since \texttt{GPT-4o} and \texttt{DeepSeek-R1} can return more than five relevant modules per query, we also set a maximum number of correctly recalled modules for a fair comparison. Concretely, for a set of $Q$ queries, recall is computed as follows: 
$$\text{Recall} = \frac{\sum_{q \in Q} \min(5, \left| \text{Retrieved}_q \cap \text{Relevant}_q \right|)}{\sum_{q \in Q} \min\left(5,\ \left| \text{Relevant}_q \right|\right)}
$$
where $\text{Retrieved}_q$ is the set of modules that the pipeline predicts as being relevant and $\text{Relevant}_q$ is the set of modules that annotators labeled as relevant for a given query $q$.

As reported in Sec.~\ref{sec:recall}, we compare the recall of our pipeline against using either \texttt{GPT-4o} or \texttt{DeepSeek-R1} as a zero-shot classifier. The prompt we use is as follows: 

\aptLtoX[graphic=no,type=html]{}{\begin{quote}}
\begin{verbatim}
You are a helpful assistant. Your task is to
determine whether a document is relevant for
answering a given query. 

Return 1 if the document is RELEVANT and 0 if
the document is NOT RELEVANT. Provide only 0 or 1.
Do not provide any additional information.
\end{verbatim}
\aptLtoX[graphic=no,type=html]{}{\end{quote}}

\paragraph{Ablations on Router Design} Next, we run additional ablations on the router design, comparing the retrieval-rerank router with a retrieval only version and a baseline \texttt{GPT-4o} model. In Retrieve Only, we use cosine similarity with embeddings to retrieve modules and select all modules with similarity scores greater than a threshold $\tau = 0.3$. We evaluate Answer Relevance and Context Relevance on a pilot dataset consisting of 100 queries that were annotated by the first author. As shown in Table~\ref{tab:ablations}, the two-step Retrieve-Rerank yields the highest answer relevancy of $89.0\%$, although the Retrieve Only ablation reports better context relevance. 

\paragraph{Embedding Model Comparisons} Finally, we evaluate different embedding models used during the retrieval process. In total, we benchmark five models: three open-source Sentence Transformers~\cite{reimers2019sentence} (\texttt{all-MiniLM-L6-v2}, \texttt{all-mpnet-base-v2}, \texttt{msmarco-distilbert-\newline cos-v5}) and two commercial models (OpenAI's \texttt{text-embedding-3-\newline small}~\cite{openaiembed} and Voyage AI's \texttt{voyage-3-lite}). As shown in Table~\ref{tab:embeddings}, \texttt{voyage-3-lite} is the most performant with a recall of $0.76$. Compared to others methods that report similar recall scores (\texttt{text-\newline embedding-3-small} and \texttt{msmarco-distilbert-cos-v5}),\newline \texttt{voyage-3-lite} has a lower latency of 399ms.

\begin{table}[b]
    \centering
    \begin{tabular}{lrrr}
    \toprule
         &  \makecell{Answer\\Relevance} & \makecell{Relevance\\@ 1} & \makecell{Relevance\\@ 5}\\
    \midrule
         Baseline & 63.0\%  & - & -\\ 
         Retrieve Only & 84.0\%  & \textbf{0.59} & \textbf{0.61}\\
         Retrieve-Rerank & \textbf{89.0\%} & 0.47 & 0.41\\
    \bottomrule
    \end{tabular}
    \Description[Router comparison table]{Comparison of router designs showing performance metrics across different approaches. The two-stage Retrieve-Rerank pipeline outperforms a Retrieve Only router and the baseline model in terms of answer relevance and recall, though with slightly lower context relevance scores.}
    \caption{We compare our two-stage pipeline with a retrieve-only router and the baseline \texttt{GPT-4o} model. We report answer and context relevancy on a pilot test set of 100 queries.}
    \label{tab:ablations}
\end{table}
\begin{table}[]
    \centering
    \begin{tabular}{lrr}
    \toprule 
    Embedding Type &  Recall & Latency\\
    \midrule
    \texttt{all-MiniLM-L6-v2} & 0.58 & \textbf{162ms}\\ 
    \texttt{all-mpnet-base-v2} & 0.67 & 592ms \\
    \texttt{msmarco-distilbert-cos-v5} & 0.74 & 623ms\\
    \texttt{text-embedding-3-small} & 0.74 & 654ms\\
    \texttt{voyage-3-lite} & \textbf{0.75} & 399ms \\ 
    \bottomrule 
    \end{tabular}
    \Description[Embedding model comparison table]{Comparison of embedding models for the retrieval process, showing recall performance and latency metrics. Five models are evaluated (all-MiniLM-L6-v2, all-mpnet-base-v2, msmarco-distilbert-cos-v5, text-embedding-3-small, and voyage-3-lite), with voyage-3-lite offering the best balance of high recall (0.76) and reasonable latency (399ms).}
    \caption{\texttt{voyage-3-lite} embeddings offer both high recall and low latency. We compare different embedding models on module recall over our test set of 100 queries and average latency (ms) over five prompts sent to ChatGPT.}
    \label{tab:embeddings}
\end{table}

\section{Query Clustering}
\label{sec:app_query}
In Sec.~\ref{sec:cluster}, we report on the types of queries that users sent to Knoll during the public deployment. To obtain the high-level clustering of query types, we used the following pipeline proposed by Tamkin et al.~\cite{tamkin2024clio}. 

First, we use \texttt{GPT-4o} to extract the task that the model is being asked to perform for each query. We use the following prompt: 

\aptLtoX[graphic=no,type=html]{}{\begin{quote}}
\begin{verbatim}
What task is the model being asked to perform
in this message?

Return just the task in the format "The task
the model is being asked to perform is [TASK]."
Do NOT provide any other information.
\end{verbatim}
\aptLtoX[graphic=no,type=html]{}{\end{quote}}

Then, we extract the embeddings for each of the high-level tasks using the \texttt{all-MiniLM-L6-v2} model. We group the embeddings using $k$-means clustering, resulting in 39 clusters. For each cluster, we then generate a title and short two-sentence description, providing all of the extracted tasks as input to an LLM. We use the following prompt adapted from Tamkin et al.~\cite{tamkin2024clio}: 

\aptLtoX[graphic=no,type=html]{}{\begin{quote}}
\begin{verbatim}
You are tasked with summarizing a group of
related statements into a short, precise, and
accurate description and name. Your goal is
to create a concise summary that captures the
essence of these statements and distinguishes
them from other similar groups of statements.
    
Summarize all the statements into a clear, precise,
two-sentence description in the past tense.
Your summary should be specific to this group. 
    
After creating the summary, generate a short name
for the group of statements. This name should be
at most ten words long (perhaps less) and be specific
but also reflective of most of the statements
(rather than reflecting only one or two). Be as
descriptive as possible and assume neither good nor
bad faith. Do not hesitate to identify and describe
socially harmful or sensitive topics specifically;
specificity is necessary for monitoring.

Present your output in the following JSON format:
{
    "summary": 
        "[Insert your two-sentence summary here]",
    "name": 
        "[Insert your generated short name here]"
}
\end{verbatim}
\aptLtoX[graphic=no,type=html]{}{\end{quote}}

Finally, we manually de-duplicated clusters. For example, the low-level of clusters of ``Spring Course Selection Assistance,'' `` Student Career and Activity Guidance'', and ``Educational Guidance and CS Course Optimization' were grouped into the ``Education \& Career Development Assistance'' category. After de-duplication, we report the 16 groups listed in the main body. 

\section{User Evaluation}
\label{sec:app_user_eval}
\begin{figure}
    \centering
    \includegraphics[width=\linewidth]{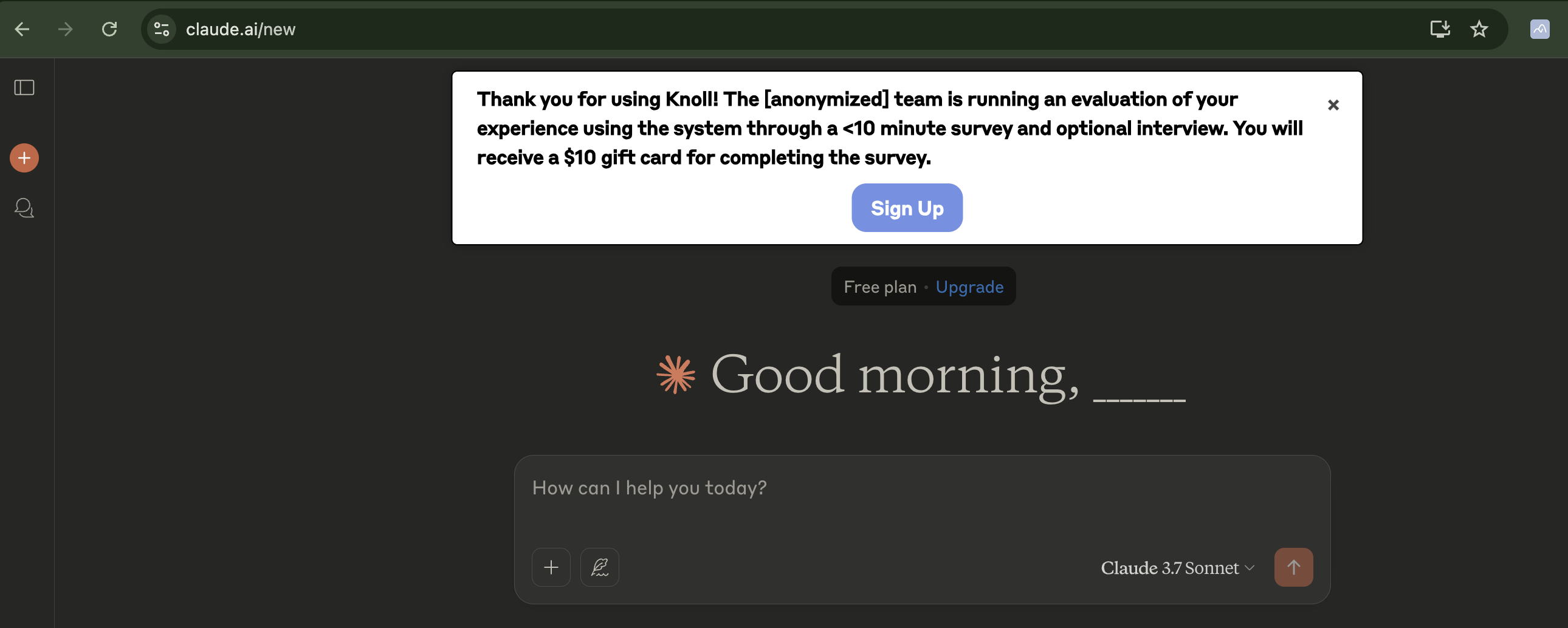}
    \Description[Claude interface with recruitment banner]{Claude's "New Chat" interface showing a recruitment banner at the top inviting users to provide feedback in a <10 minute survey. Users will receive a \$10 gift card for completing the survey.}
    \caption{Users were shown a banner on the ChatGPT and Claude homepage, recruiting them to provide feedback on their experience, after one week of using Knoll}
    \label{fig:recruitment}
\end{figure}
In total, 203 users downloaded Knoll in the two months since deploying our prototype. We provide details on the 30 predefined modules that users added to their extension and the nine modules that users created in Table~\ref{tab:modules}.  

We collect qualitative survey data from 17 users and follow-up interviews with eight (see Table~\ref{tab:participants}). Fig.~\ref{fig:recruitment} shows the advertisement banner used to recruit participants. Below, we also have listed the questions we asked in our survey and interviews. 

\subsection{Survey Questions}
\begin{enumerate}
    \item What types of queries did you send when using Knoll that you otherwise would not have asked ChatGPT or Claude?
    \item What was the best part about using Knoll? What was the worst part?
    \item What other types of external knowledge would you find useful for AI chatbots to have access to?
\end{enumerate}

Survey participants were then shown queries they sent with Knoll enabled and two responses --- one generated with Knoll modules inserted and the other from a baseline \texttt{GPT-4o} model. For each query, we ask them to select which response they preferred. 

\subsubsection{Interview Questions}
\begin{enumerate}
    \item Could you briefly describe how you use ChatGPT / Claude in your day-to-day life? 
    \item How do you typically add contextualized knowledge to ChatGPT / Claude (e.g., via prompting, using CustomGPTs)? How do you find this process?
    \item What types of queries did you send to ChatGPT / Claude using Knoll? 
    \item What was your intent when sending this prompt to ChatGPT / Claude? 
    \item When sending this prompt, what kind of response were you looking for? 
    \item Did you add any custom knowledge modules to Knoll? If so, what type of knowledge did you add?
    \item Were there times when you think having Knoll installed made the response from ChatGPT / Claude more helpful?
    \item Were there times when you think having Knoll installed made the response from ChatGPT / Claude less helpful?
    \item What other types of information or knowledge do you think would be helpful for ChatGPT / Claude to have access to? 
\end{enumerate}

\begin{table*}[]
    \centering
    \begin{tabular}{lrrrr}
    \toprule
    \normalsize
    \textbf{Module Name} & 
    \normalsize
    \textbf{\# of Users} &
    \normalsize
    \textbf{Token Count} & 
    \normalsize
    \textbf{\# of Queries} & 
    \normalsize
    \textbf{User Discovered}\\
    \midrule

\emph{\textbf{Predefined Modules}}\\
Welcome to Knoll & 198 & 899 & 20 \\
CS Summer Internship Postings 2025 & 49 & 10953 & 214 \\
CHI 2025 Papers & 22 & 250329 & 161 \\
AI Track Requirements & 17 & 21856 & 9 \\
Summer Research Application Tips & 15 & 860 & 44 \\
CS Courses & 10 & 20002 & 20 \\
CS Course Reviews  & 10 & 10849 & 34 & \cmark\\
CHI 2025 Program & 9 & 71797 & 51 \\
CHI 2025 HCI + AI Papers & 6 & 49743 & 17 \\
Reasoning LLMs & 6 & 2546 & 52 & \cmark\\
Other Summer Internship Postings 2025 & 6 & 5587 & 27 \\
CS PhD Course Requirements & 5 & 919 & 10 \\
Course Study Guide & 3 & 3935 & 8 \\
CS SoP Tips & 3 & 708 & 1 & \cmark \\
Research Elevator Pitches & 3 & 426 & 1 & \cmark\\
Awesome RAG Papers & 3 & 6907 & 20 & \cmark \\
HCI Track Requirements & 3 & 24591 & 54 \\
ICLR 2025 Program & 2 & 387867 & 9 & \cmark \\
Computer Graphics Track Requirements & 2 & 20655 & 4 \\
Advice on Being a CS PhD & 2 & 8301 & 3 & \cmark \\
Yokohama Recommendations & 1 & 11820 & 17 & \cmark \\
Unspecialized Track Requirements & 1 & 23637 & 5 \\
Math Courses & 1 & 7238 & 3 & \cmark \\
Design Courses & 1 & 4479 & 0 & \cmark \\
Comp. Bio Track Requirements & 1 & 14158 & 19 \\
Mathematical \& CS Courses  & 1 & 228 & 5 & \cmark \\
Theory Track Requirements & 1 & 24240 & 1 \\
Spring Courses for CS PhDs & 1 & 7313 & 0 \\
CS AI-Related Courses & 1 & 1256 & 1 &  \cmark\\
Course Assignment Writing Guidelines & 1 & 1343 & 16 &  \cmark \\
\hdashline
\emph{\textbf{User-Created Modules}}\\
    Personal Module & 8 & - & 150 & \cmark\\
    Paper Intro Writing & 1 & 2527 & 3 & \cmark\\
    Course Notes & 1 & 14360 & 1 &  \cmark\\ 
    Course Exam Prep & 1 & 14558 & 1 &  \cmark\\
    Job Descriptions & 1 &1998 & 0 &  \cmark\\
    Personal Notes & 1 & - & 1 &  \cmark\\ 
    Community Policies & 1 & - & 2 &  \cmark\\
    Exam \& HW FAQ & 1 & - & 15 &  \cmark\\
    Incomplete Policies & 1 & - & 15 &  \cmark\\
    Course Policies & 1 & - & 15 &  \cmark\\
    \bottomrule 
         & 
    \end{tabular}
    \Description[Detailed table of all knowledge modules]{Detailed table listing all knowledge modules in the study, showing module name, number of users, token count, number of queries, and whether the module was user-discovered. The table is divided into thirty "Predefined Modules" and ten "User-Created Modules" sections.}
    \caption{Users imported a diverse range of pre-defined knowledge modules in addition to contributing their own modules to the ecosystem. We report the total numbers of users who added the module, module size (in tokens), total number of queries sent with the module activated, and whether the module was advertised to users or organically discovered. We do not report the token count for user-created modules that are private and do not grant the authors' access privileges.}
    \label{tab:modules}
\end{table*}

\begin{table*}[]
    \centering
    \begin{tabular}{llll>{\raggedright\arraybackslash}p{3.25in}}
    \toprule
    \textbf{ID} &  \textbf{\makecell{Follow-Up\\Interview}} & \textbf{\makecell{Queries with\\$\geq 1$ Module}} & \makecell{\textbf{Total}\\\textbf{Queries}} & \textbf{Added Modules} \\
    \midrule
    P1 & \checkmark & 56 & 104 & CS Summer Internship Postings 2025, Summer Research Application Tips\\
    P2 & & 10 & 624 & CS Summer Internship Postings 2025\\
    P3 & \checkmark & 27 & 660 & Course Assignment Writing Guidelines, Paper Intro Writing, CS Summer Internship Postings 2025, CHI 2025 Program, CHI 2025 Papers\\
    P4 & & 2 & 5 & CS Summer Internship Postings 2025\\
    P5 & & 54 & 964 & CS Summer Internship Postings 2025\\
    P6 & \checkmark & 48 & 204 & CS Course Reviews, CS AI-Related Courses, CS Summer Internship Postings 2025 \\ 
    P7 & & 28 & 240 & AI Track Requirements, Design Courses, HCI Track Requirements\\
    P8 & &3 & 180 & AI Track Requirements\\
    P9 & \checkmark& 100 & 132 & CHI 2025 Program, CHI 2025 Program, Job Descriptions\\
    P10 & & 10 & 93 & Personal Module\\
    P11 & \checkmark &21 & 541 & Comp. Bio Track Requirements\\ 
    P12 & & 3 & 125 & CHI 2025 Papers\\
    P13 & & 6 & 271 & CS Courses, AI Track Requirements\\ 
    P14 & \checkmark & 2 & 3 & CS Summer Internship Postings 2025, Summer Research Application Tips\\
    P15 & \checkmark & 4 & 4 & Course Study Guide\\ 
    P16 & & 23 &48 & Course Study Guide\\
    P17 & \checkmark & 5 & 44 & CS Courses, AI Track Requirements\\ 
    \bottomrule 
    
    \end{tabular}
    \Description[Survey participant information table]{Survey participant information table showing participant IDs, whether they participated in follow-up interviews, total messages sent, and which modules they added to their Knoll extension.}
    \caption{Information on our 17 survey respondents including number of queries where at least one module was activated, total number of queries sent, modules added, and whether they participated in a follow-up interview.}
    \label{tab:participants}
\end{table*}

%TC:endignore

\end{document}